\documentclass[useAMS,usenatbib]{mn2e}

\usepackage{epsfig}
\usepackage{amsmath}
\usepackage{amssymb}
\usepackage{mathrsfs}
\usepackage{graphicx}
\usepackage{relsize}
\usepackage{mathtools}
\usepackage{multirow}
\usepackage{hyperref}

\hypersetup{
     colorlinks   = true,
     citecolor    = blue
}

\newcommand{\apj}{ApJ} 
\newcommand{\apjl}{ApJ} 
\newcommand{\apjs}{ApJ} 
\newcommand{\aap}{A\&A} 
\newcommand{\araa}{ARAA} 
\newcommand{\mnras}{MNRAS} 
\newcommand{\nat}{Nature} 
\newcommand\aapr{A\&A~Rev.} 

\newcommand{\rflat}{R_\mathrm{flat}}          
\newcommand{\wfil}{W_\mathrm{fil}}
\newcommand{\alphavir}{\alpha_\mathrm{vir}}

\newcommand{\mach}{\mathcal{M}}
\newcommand{\macha}{\mathcal{M}_\mathrm{A}}

\newcommand{\kinj}{k_\mathrm{inj}}

\newcommand{\msol}{\mbox{$M_{\sun}$}}

\newcommand{\tff}{t_\mathrm{ff}}

\newcommand{\cs}{c_\mathrm{s}}

\newcommand{\km}{\mathrm{km}}
\newcommand{\pc}{\mathrm{pc}}
\newcommand{\AU}{\mbox{AU}}
\newcommand{\s}{\mathrm{s}}

\newcommand{\Myr}{\mathrm{Myr}}
\newcommand{\Gauss}{\mathrm{G}}
\newcommand{\cm}{\mbox{cm}}
\newcommand{\g}{\mbox{g}}
\newcommand{\G}{\mbox{G}}

\title[Universality of interstellar filaments]{On the universality of interstellar filaments: theory meets simulations and observations}

\author[Federrath]{
Christoph~Federrath$^{1}$\thanks{E-mail: christoph.federrath@anu.edu.au}\\
$^{1}$Research School of Astronomy and Astrophysics, Australian National University, Canberra, ACT~2611, Australia
}

\begin{document}

\maketitle

\begin{abstract}
Filaments are ubiquitous in the universe. Recent observations have revealed that stars and star clusters form preferentially along dense filaments. Understanding the formation and properties of filaments is therefore a crucial step in understanding star formation. Here we perform three-dimensional high-resolution magnetohydrodynamical simulations that follow the evolution of molecular clouds and the formation of filaments and stars. We apply a filament detection algorithm and compare simulations with different combinations of physical ingredients: gravity, turbulence, magnetic fields and jet/outflow feedback. We find that gravity-only simulations produce significantly narrower filament profiles than observed, while simulations that include turbulence produce realistic filament properties. For these turbulence simulations, we find a remarkably universal filament width of $0.10\pm0.02\,\pc$, which is independent of the star formation history of the clouds. We derive a theoretical model that provides a physical explanation for this characteristic filament width, based on the sonic scale ($\lambda_\mathrm{sonic}$) of molecular cloud turbulence. Our derivation provides $\lambda_\mathrm{sonic}$ as a function of the cloud diameter $L$, the velocity dispersion $\sigma_v$, the gas sound speed $\cs$, and the ratio of thermal to magnetic pressure, plasma $\beta$. For typical cloud conditions in the Milky Way spiral arms, we find \mbox{$\lambda_\mathrm{sonic}=0.04$--$0.16\,\pc$}, in excellent agreement with the filament width of \mbox{$0.05$--$0.15\,\pc$} from observations. Consistent with the theoretical model assumptions, we find that the velocity dispersion inside the filaments is subsonic and supersonic outside. We further explain the observed $p=2$ scaling of the filament density profile, $\rho\propto r^{-p}$ with the collision of two planar shocks forming a filament at their intersection.
\end{abstract}


\section{Introduction}

Interstellar filaments have recently attracted much attention, especially since the Herschel satellite revealed a wealth of filamentary structures in both star-forming and quiescent clouds \citep{AndreEtAl2010,MenshchikovEtAl2010,MivilleDeschenesEtAl2010,ArzoumanianEtAl2011,HillEtAl2011,RoyEtAl2015}. It is believed that these filaments are fundamental building blocks of molecular clouds and that they must play an important role for star formation \citep{SchneiderElmegreen1979,BalsaraEtAl2001,AndreEtAl2014}. This is because the dense gas is organized in filamentary structures. Star-forming cores appear primarily along dense filaments \citep{KonyvesEtAl2015}, with young star clusters being located at their intersections \citep{Myers2011,SchneiderEtAl2012}.

A key property obtained from recent observations and simulations of interstellar filaments is that they seem to have an almost universal characteristic width of about $0.1\,\pc$ \citep{ArzoumanianEtAl2011,PalmeirimEtAl2013,JuvelaEtAl2012a,MalinenEtAl2012,BenedettiniEtAl2015,KirkEtAl2015,KainulainenEtAl2015}, coherent velocity structures \citep{HacarEtAl2013,MoeckelBurkert2015,SmithEtAl2016} and orientations preferentially (but not always) perpendicular to the magnetic field direction \citep{SugitaniEtAl2011,GaenslerEtAl2011,PalmeirimEtAl2013,Hennebelle2013,Tomisaka2014,PlanckMagneticFilaments2014,PillaiEtAl2015,PlanckMagneticFilaments2015_1,PlanckMagneticFilaments2015_2,SeifriedWalch2015}. Our goal here is to unravel the origin of the universality of filaments and to provide a physical explanation for the characteristic filament width of $\sim0.1\,\pc$ found in observations.

In order to make progress and contribute to our understanding of interstellar filaments, we use numerical simulations that follow the dynamical evolution of molecular clouds and the formation of stars within them. We compare six simulation models with different combinations of physical ingredients to evaluate their individual roles and combined effects: gravity, turbulence, magnetic fields, and jet/outflow feedback. We analyse the filaments profiles for each simulation model and compare them to observations. We find that only the models that at least have turbulence included produce filament distributions and widths consistent with observations, while filaments in the gravity-only models are significantly narrower by at least a factor of 2.

Based on the finding that only models with turbulence produce realistic filament widths, we provide a physical explanation for the observed characteristic width of interstellar filaments. In this theoretical model, the sonic scale of the turbulence provides the natural scale of filament formation in shocks. We derive the sonic scale and its dependences on the cloud size, the velocity dispersion, the sound speed and the magnetic field strength in the cloud. We find the theoretical prediction of \mbox{$\lambda_\mathrm{sonic}=0.04$--$0.16\,\pc$}, which is in excellent agreement with the observed range of filament widths, \mbox{$0.05$--$0.15\,\pc$}.

The paper is organized as follows. First, we summarize our simulation techniques and the filament detection algorithm in Section~\ref{sec:methods}. We then present column density projections of our simulation models with detected filaments highlighted and radial filament profiles analysed, as well as a direct comparison with observations in Section~\ref{sec:results}. Our theoretical model for the university of filament widths based on the sonic scale of molecular cloud turbulence is presented in Section~\ref{sec:physics}. Finally, we summarize and conclude in Section~\ref{sec:conclusions}.

\section{Simulation and analysis methods} \label{sec:methods}

\subsection{Numerical simulations}

We use the multi-physics, adaptive mesh refinement \citep[AMR][]{BergerColella1989} code \textsc{flash} \citep{FryxellEtAl2000,DubeyEtAl2008} in its latest version~(v4), to solve the compressible magnetohydrodynamical (MHD) equations on three-dimensional (3D) periodic grids of fixed side length $L$, including turbulence, magnetic fields, self-gravity and outflow feedback. The positive-definite HLL3R Riemann solver \citep{WaaganFederrathKlingenberg2011} is used to guarantee stability and accuracy of the numerical solution of the MHD equations.

\subsubsection{Turbulence driving} \label{sec:turbdriving}
Turbulence is a key for star formation \citep{MacLowKlessen2004,ElmegreenScalo2004,McKeeOstriker2007,HennebelleFalgarone2012,Hopkins2013IMF,Krumholz2014,PadoanEtAl2014}, so most of our simulations include a turbulence driving module\footnote{Note that molecular cloud turbulence is likely driven by a combination of supernova explosions, stellar feedback in the form of jets, outflows, winds, radiation fronts and shells, gravitational contraction, magneto-rotational instability and galactic spiral-arm compression. Since turbulence decays very quickly \citep{MacLowEtAl1998,StoneOstrikerGammie1998} and turbulence is observed on all spatial scales in the ISM, the turbulence must be driven, which is modelled with the turbulence driving procedure explained here.} that produces turbulence similar to what is observed in real molecular clouds, i.e., driving on the largest scales \citep{HeyerWilliamsBrunt2006,BruntHeyerMacLow2009} and with a power spectrum, $E(k)\sim k^{-2}$, consistent with supersonic, compressible turbulence \citep{Larson1981,HeyerBrunt2004,RomanDuvalEtAl2011}. This type of turbulence spectrum is consistent with simulations of supersonic turbulence \citep{KritsukEtAl2007,SchmidtEtAl2009,FederrathDuvalKlessenSchmidtMacLow2010,Federrath2013}. We drive turbulence by applying a stochastic Ornstein-Uhlenbeck process \citep{EswaranPope1988,SchmidtHillebrandtNiemeyer2006} to construct an acceleration field ${\mathbfit{F}_\mathrm{stir}}$, which serves as a momentum and energy source term in the momentum equation. As suggested by observations, ${\mathbfit{F}_\mathrm{stir}}$ only contains large-scale modes, $1<\left|\mathbf{k}\right|L/2\pi<3$, where most of the power is injected at the $\kinj=2$ mode in Fourier space, i.e., on half of the size of the computational domain. The turbulence on smaller scales is not directly affected by the driving and develops self-consistently from the cascade of energy originating on larger scales. The turbulence forcing module used here excites a natural mixture of solenoidal and compressible modes, corresponding to a turbulent driving parameter $b=0.4$ \citep{FederrathDuvalKlessenSchmidtMacLow2010}, although some cloud-to-cloud variations in this parameter from $b\sim1/3$ (purely solenoidal driving) to $b\sim1$ (purely compressive driving) are expected for real clouds \citep{PriceFederrathBrunt2011,GinsburgFederrathDarling2013,KainulainenFederrathHenning2013}.

\begin{table*}
\caption{Key simulation parameters and detected filament properties.}
\label{tab:sims}
\def\arraystretch{1.1}
\setlength{\tabcolsep}{2.9pt}
\begin{tabular}{lccccccccccc}
\hline
Simulation Model & Gravity & Turbulence & $B (\mu\Gauss)$  & Jets & $\sigma_v (\km/\s)$ & $\mach$ & $\beta$ & $\macha$ & $N_\mathrm{res}^3$ & $N_\mathrm{filaments}$ & Width $\wfil\,(\pc)$ \\
(1) & (2) & (3) & (4) & (5) & (6) & (7) & (8) & (9) & (10) & (11) & (12) \\
\hline
1.~Gravity only (Gauss ICs)  & Yes &  None & $\infty$  & No &    $0$ & $0$    & $0$      & $\infty$ & $1024^3$ & $17$ & $0.04\pm0.01$ \\
2.~Gravity only (Turb ICs)     & Yes &  None & $\infty$  & No &    $0$ & $0$    & $0$      & $\infty$ & $1024^3$ & $37$ & $0.05\pm0.02$ \\
3.~Turbulence+Magnetic Fields    & No  &  Mix    & $10$      & No & $1.0$ & $5.0$ & $0.33$ & $2.0$    & $1024^3$ & $36$ & $0.10\pm0.02$ \\
4.~Gravity vs.~Turbulence            & Yes &  Mix    & $\infty$  & No & $1.0$ & $5.0$ & $0$      & $\infty$  & $1024^3$ & $29$ & $0.10\pm0.03$ \\
5.~Grav vs.~Turb+Magnetic         & Yes &  Mix    & $10$     & No  & $1.0$ & $5.0$ & $0.33$ & $2.0$    & $1024^3$ & $33$ & $0.08\pm0.02$ \\
6.~Grav vs.~Turb+Mag+Jets       & Yes &  Mix    & $10$     & Yes & $1.0$ & $5.0$ & $0.33$ & $2.0$    & $1024^3$ & $33$ & $0.10\pm0.02$ \\
\hline
7.~Grav vs.~Turb+Mag+Jets ($512^3$) & Yes & Mix & $10$ & Yes & $1.0$ & $5.0$ & $0.33$ & $2.0$ & $512^3$ & $26$ & $0.09\pm0.02$ \\
\hline
\end{tabular}
\begin{minipage}{\linewidth}
\textbf{Notes.} Column 1: simulation model. The first two models are gravity-only simulations, one from Gaussian initial conditions (ICs) and the other from turbulent ICs. The third model is a pure MHD turbulence simulation (no gravity, no star formation). The fourth, fifth and sixth models are all star formation simulations including gravity, with increasing complexity and number of physical processes (turbulence, then adding magnetic fields, and finally also adding jet/outflow feedback). The last model is identical to the previous one, but was run with a lower grid resolution to check numerical convergence. Columns 2--5: whether gravity and star formation are included, the type of turbulence driving \citep{FederrathDuvalKlessenSchmidtMacLow2010,Federrath2013}, the magnetic field strength, and whether jet/outflow feedback is included. Columns 6 and 7: turbulent velocity dispersion, and turbulent rms sonic Mach number. Columns 8 and 9: ratio of thermal to magnetic pressure (plasma $\beta$), and Alfv\'en Mach number. Column 10: maximum effective grid resolution (note that refinement is based on the Jeans length with a minimum of 32 cells per Jeans length). Columns 11 and 12: number of detected filaments ($N_\mathrm{filaments}$) and derived width ($\wfil$) of the filaments.
\end{minipage}
\end{table*}

\subsubsection{Sink particles} \label{sec:sinks}
In order to follow star formation and gas accretion, we use the sink particle method developed by \citet{FederrathBanerjeeClarkKlessen2010}. Sink particles form dynamically in our simulations when a local region in the simulation domain undergoes gravitational collapse and forms stars. This is technically achieved by first flagging each computational cell that exceeds the Jeans resolution density,
\begin{equation}
\rho_\mathrm{sink} = \frac{\pi\cs^2}{G\lambda_\mathrm{J}^2} = \frac{\pi\cs^2}{4 G r_\mathrm{sink}^2},
\end{equation}
where $\cs$ is the sound speed, $G$ is the gravitational constant, and $\lambda_\mathrm{J}$ is the local Jeans length. This determines the sink particle accretion radius, $r_\mathrm{sink} = \lambda_\mathrm{J}/2$, which is set to $2.5$ grid cell lengths in order to capture star formation and to avoid artificial fragmentation on the highest level of AMR \citep{TrueloveEtAl1997}. If the gas density in a cell exceeds $\rho_\mathrm{sink}$, a spherical control volume with radius $r_\mathrm{sink}$ is constructed around the cell and it is checked that all the gas within the control volume is Jeans-unstable, gravitationally bound and collapsing towards the central cell. A sink particle is only formed in the central cell of the control volume, if all of these checks are passed. This avoids spurious formation of sink particles and guarantees that only bound and collapsing gas forms stars \citep{FederrathBanerjeeClarkKlessen2010}.

On all the lower levels of AMR (except the highest level, where sink particles form), we use an adaptive grid refinement criterion based on the local Jeans length, such that $\lambda_\mathrm{J}$ is always resolved with at least 32 grid cell lengths in each of the three spatial directions of our 3D domain. This resolution criterion is very conservative and computationally costly, but guarantees that we resolve turbulence on the Jeans scale \citep{FederrathSurSchleicherBanerjeeKlessen2011}, potential dynamo amplification of the magnetic field in the dense cores \citep{SurEtAl2010}, and capture the basic structure of accretion discs forming on the smallest scales \citep{FederrathEtAl2014}. If a cell within the accretion radius of an existing sink particle exceeds $\rho_\mathrm{sink}$ during the further evolution, is bound to the sink particle and is moving towards it, then we accrete the excess mass above $\rho_\mathrm{sink}$ on to the sink particle, conserving mass, momentum and angular momentum.  We compute all contributions to the gravitational interactions between the gas on the grid \citep[with the iterative multigrid solver by][]{Ricker2008} and the sink particles (by direct summation over all sink particles and grid cells). A second-order leapfrog integrator is used to advance the sink particles on a timestep that allows us to resolve close and highly eccentric orbits \citep[for details, see the tests in][]{FederrathBanerjeeClarkKlessen2010}.

\subsubsection{Outflow/Jet feedback} \label{sec:jets}
Powerful jets and outflows are launched from the protostellar accretion discs around newborn stars. These outflows carry enough mass, linear and angular momentum to transform the structure of their parent molecular cloud and to potentially control star formation in a feedback loop \citep{Federrath2015}. In order to take this most important mechanical feedback effect \citep{KrumholzEtAl2014} into account, we recently extended the sink particle approach such that sink particles can launch fast collimated jets together with a wide-angle, lower-speed outflow component, to reproduce the global features of observed jets and outflows, as well as to be consistent with high-resolution simulations of the jet launching process and with theoretical predictions \citep{FederrathEtAl2014}. The most important feature of our jet/outflow feedback model is that it converges and reproduces the large-scale effects of jets and outflows with relatively low resolution, such as with sink particle radii of $r_\mathrm{sink}\sim1000\,\AU$, used here. Our feedback module has been carefully tested and compared to previous implementations of jet/outflow feedback such as the models implemented in \citet{WangEtAl2010} and \citet{CunninghamEtAl2011}. The most important difference to any previous implementation is that our feedback model includes angular momentum transfer, reproduces the fast collimated jet component and demonstrated convergence \citep[for details, see][]{FederrathEtAl2014}.

\subsubsection{Simulation parameters} \label{sec:simparams}
All our simulations share the same global properties: a cloud size $L=2\,\pc$, a total cloud mass $M=388\,\msol$ and a mean density $\rho_0=3.28\times10^{-21}\,\g\,\cm^{-3}$, resulting in a global mean freefall time $\tff=1.16\,\Myr$. Models including turbulence have a velocity dispersion $\sigma_v=1\,\km\,\s^{-1}$ and an rms Mach number of $\mach=5$, maintained by the turbulence driving (Sec.~\ref{sec:turbdriving}). We use a fixed sound speed $\cs=0.2\,\km\,\s^{-1}$, appropriate for molecular gas with temperature $T=10\,\mathrm{K}$ over the wide range of densities that lead to filament and dense core formation \citep{OmukaiEtAl2005}. Finally, models including a magnetic field start with a uniform initial field of $B=10\,\mu\G$, which is subsequently compressed, tangled and twisted by the turbulence, similar to how it would be structured in real molecular clouds. The magnetic field strength, the turbulent velocity dispersion and the mean density all follow typical values derived from observations of clouds with the given physical properties \citep{FalgaronePugetPerault1992,CrutcherEtAl2010}. This leads to the dimensionless virial ratio $\alphavir=1.0$ \citep[also typical for molecular clouds in the Milky Way; see][]{FalgaronePugetPerault1992,KauffmannEtAl2013,HernandezTan2015} and to a plasma beta parameter (ratio of thermal to magnetic pressure) $\beta=0.33$ or an Alfv\'en Mach number $\macha=2.0$. \citet{FalgaroneEtAl2008} find an average Alfv\'en Mach number of about $\macha=1.5$ in 14 different star-forming regions in the Milky Way. Thus, the assumed magnetic field in our simulation models is close to the values typically observed in molecular clouds and in cloud cores.

We run six basic models, which---step by step---include more physics (see Table~\ref{tab:sims}). In the first two simulations we only include self-gravity. The first one (`Gravity only (Gauss ICs)') uses Gaussian initial density perturbations, while the second one (`Gravity only (Turb ICs)') uses turbulent density perturbations. Neither turbulent velocities nor magnetic fields are included in these models. The third model (`Turbulence+Magnetic Fields') does not include gravity, but instead has a typical level and mixture of molecular cloud turbulence (see~\S\ref{sec:turbdriving}) and a standard magnetic field for the given cloud size and mass. In the fourth model (`Gravity vs.~Turbulence'), we include gravity and turbulence. The fifth model (`Grav vs.~Turb+Magnetic') is identical to the fourth model, but adds magnetic fields. Finally, the sixth model (`Grav vs.~Turb+Mag+Jets') is identical to the fifth model, but additionally includes jet and outflow feedback (see~\S\ref{sec:jets}). These six basic models were all run with a maximum effective grid resolution of $1024^3$ cells. Their key parameters are listed in Table~\ref{tab:sims}.

We also run an additional model, which is identical to the sixth model (with jet/outflow feedback), but has a lower maximum effective resolution of $512^3$ cells, demonstrating numerical convergence of our filament results (see Appendix~\ref{app:resnum}). The lower-resolution model is listed in the bottom row of Table~\ref{tab:sims}.

\subsection{Filament detection and analysis} \label{sec:disperse}

Our goal is to detect and analyse filaments in exactly the same way as was done in observations, in order to provide the best possible comparison of our simulations to observations, such as the ones by \citet{ArzoumanianEtAl2011}, \citet{JuvelaEtAl2012a}, \citet{MalinenEtAl2012}, \citet{PalmeirimEtAl2013}, \citet{BenedettiniEtAl2015}, and \citet{KainulainenEtAl2015}.

In order to identify filaments in the column density maps produced from our simulations, we apply the open-source tool DisPerSE \citep{Sousbie2011,SousbieEtAl2011} as in \citet{ArzoumanianEtAl2011}. The method is based on mathematical principles of topology and traces filaments by connecting saddle points to maxima with integral lines. The only free parameter of the method is the so-called `persistence threshold', which we set to $5.3\times10^{21}\,\cm^{-2}$, which is the mean column density of our model clouds. \citet{ArzoumanianEtAl2011} set the persistence threshold to a ten times lower value than this in their observations of IC~5146. In order to check the dependence of our results on the persistence threshold, we also use the same threshold as in \citet{ArzoumanianEtAl2011} and compare it to our ten times higher standard threshold in Appendix~\ref{app:pers}. This shows that the number of filaments increases with decreasing threshold and the average column density of filaments decreases, but the filament width does not change significantly with persistence threshold.

Once the DisPerSE algorithm has identified all the pixels in our column density maps that belong to individual filaments, we compute radial profiles centred on each filament. The radial profiles are computed by selecting all pixels belonging to an individual filament and then tracing all the column density cells at a perpendicular distance $r$ to the filament. Binning the average column density and column density dispersion in the radial distance $r$ from each filament yields the filament profile. We then compare the average filament profiles for each simulation listed in Table~\ref{tab:sims}, in order to investigate the dependence of the filament width on whether only gravity or additional physics such as turbulence, magnetic fields and/or feedback are included in the model clouds.

\section{Results} \label{sec:results}

Here we present the main results of our filament analysis. We start by looking at the spatial distribution and morphology of the filaments detected in each of our six basic simulation models from Table~\ref{tab:sims}. We then build the average radial profiles of the filaments and extract their characteristic widths and column densities. Finally, we compare our simulations to observations.

\subsection{Filament structure and morphology} \label{sec:structure}

\begin{figure*}
\centerline{\includegraphics[width=0.9\linewidth]{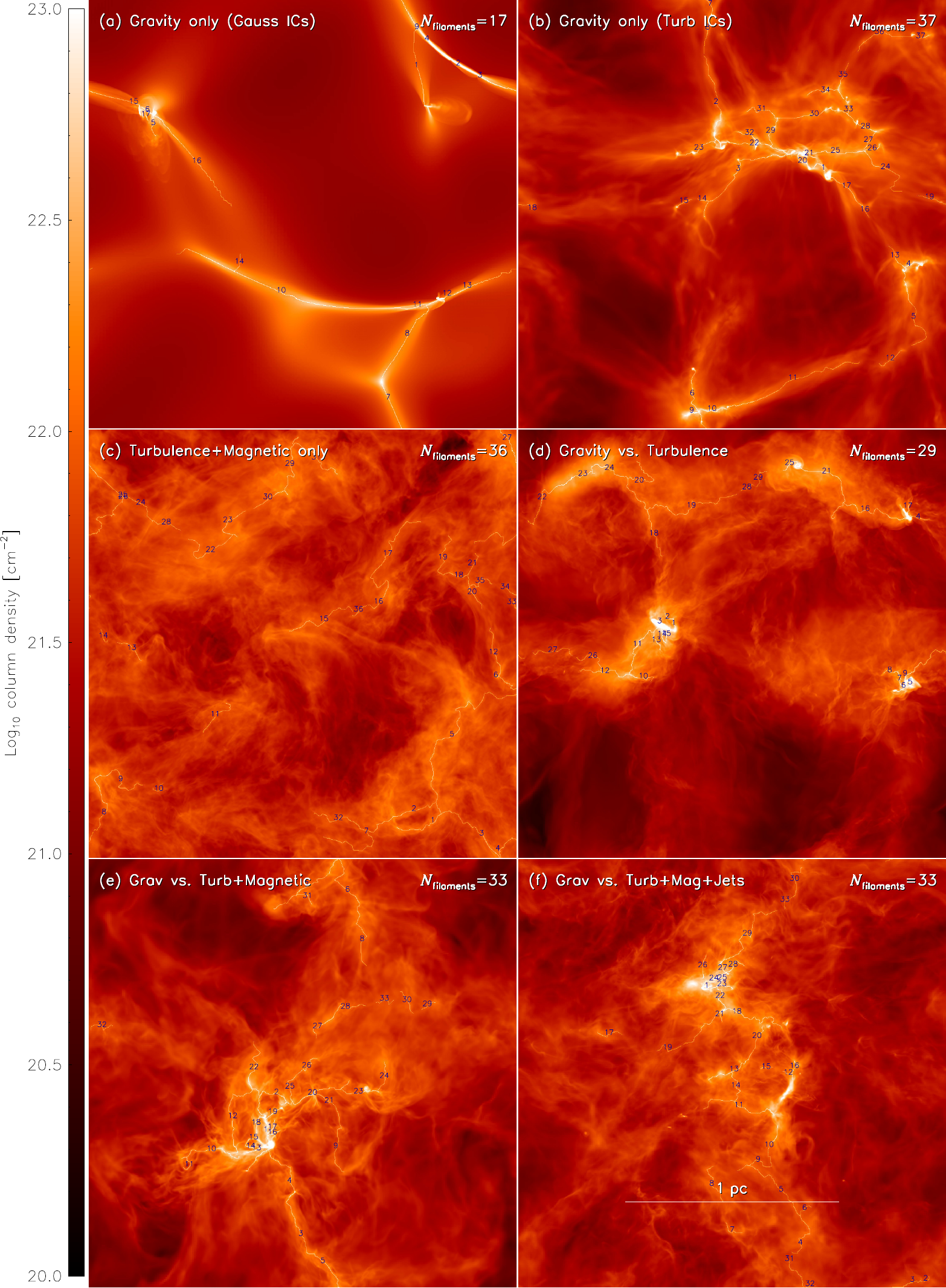}}
\caption{Column density projections with filaments highlighted and labelled in our six simulation models with different physical ingredients (see Table~\ref{tab:sims}): Gravity only from Gaussian initial conditions (ICs) (a), Gravity only from turbulent ICs (b), pure MHD turbulence (c) Gravity vs.~Turbulence (d), Gravity vs.~Turbulence + Magnetic Fields (e), and Gravity vs.~Turbulence + Magnetic Fields + Jet/Outflow Feedback (f). We see rich complex networks of filaments in all models. The respective filament profiles are computed in each of the models and for each individual filament and shown in Figure~\ref{fig:filprofs}.}
\label{fig:images}
\end{figure*}

The spatial distribution of filaments in our six basic simulations is shown in Figure~\ref{fig:images}. We distinguish simulation models with different physical ingredients and combinations of these: gravity, turbulence, magnetic fields, and jet/outflow feedback (cf.~\S\ref{sec:simparams}). The detected filaments are highlighted in the column density projections of Figure~\ref{fig:images}, by increasing the column density of the pixels belonging to filaments by a factor of 3, such that they stand out. The number of detected filaments is shown in each panel and individual filaments are numbered.

As shown in Appendices~\ref{app:res} and~\ref{app:pers}, the number of detected filaments depends on the numerical resolution, the telescope beam smoothing and the DisPerSE persistence threshold. These dependences concerning the number of detected filaments are expected, because small-scale filaments will be lost by smoothing as the numerical or telescope resolution decreases, and low-column density filaments will not be detected by DisPerSE, if the persistence threshold is set to a high value. Thus, one should be cautious when interpreting absolute numbers of detected filaments and their average column density, because these depend on resolution and detection algorithm. Remarkably however, we will see below that the characteristic width of the filaments neither depends significantly on the numerical resolution nor on the telescope beam smoothing (as long as the resolution is comparable to or higher than the width of individual filaments) and it does not vary significantly with persistence threshold, as demonstrated in Appendices~\ref{app:res} and~\ref{app:pers}. We thus concentrate in the following on analysing the converged and robust physical property of the filaments, namely the filament width, obtain from stacked and individual filament profiles.

\subsection{Filament profiles} \label{sec:profiles}

\begin{figure*}
\centerline{\includegraphics[width=1.0\linewidth]{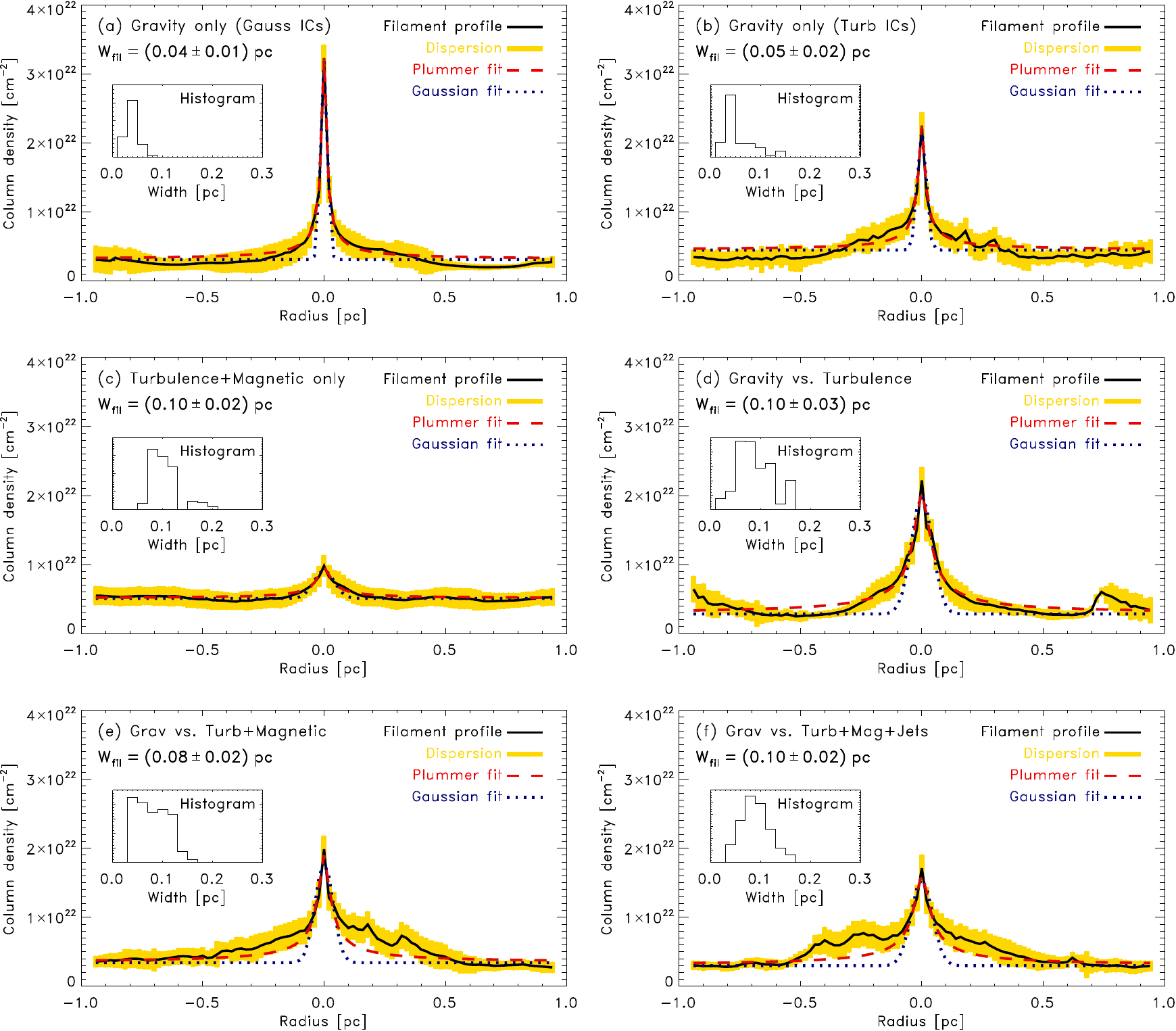}}
\caption{Radial filament profiles in the six simulation models shown in Figure~\ref{fig:images}. The filament profiles for each of the simulation models were averaged and the yellow shaded region shows the $1\sigma$-dispersion around the average filament profile (shown as solid lines). Plummer fits with Equation~(\ref{eq:plummer}) and Gaussian fits with Equation~(\ref{eq:gauss}) are shown as dashed and dotted lines, respectively. The fitted widths $\wfil$ are given in each panel and are summarized in Table~\ref{tab:sims} and the inset plots show the histograms of individual filament widths. We see that simulations that only include gravity (panels a and b) have a significantly narrower profile compared to any of the models that include turbulence (panels c--f), which are all consistent with a remarkably universal filament width of about $0.1\,\pc$.}
\label{fig:filprofs}
\end{figure*}

Filament profiles are the crucial analysis tool to determine the width of filaments. Here we compute radial profiles of each filament individually and compare them between our six main simulation models. We first select all pixels belonging to a filament and then find each pixel perpendicular to this filament at increasing radial distance. The average column density for each radial distance is recorded. We repeat this procedure for each filament and compute the average column density and $1\sigma$-dispersion around the average.

Figure~\ref{fig:filprofs} shows the filament profiles as solid lines with the yellow shaded region outlining the $1\sigma$-dispersion. As in Figure~\ref{fig:images}, we compare the six simulation models with increasing physical complexity in the same order. The filament widths are determined by two independent fits with a Plummer and a Gaussian function shown as dashed and dotted lines, respectively.

The Plummer filament profile is defined as
\begin{equation}
\Sigma(r) = \Sigma(0) \left[ 1 + \left(r/\rflat\right)^2 \right]^{(1-p)/2} + \Sigma_\mathrm{offset},
\label{eq:plummer}
\end{equation}
with the parameters $p$ and $\rflat$, where the latter is related to the filament width $W\approx3\,\rflat$ \citep{ArzoumanianEtAl2011}. We experimented with the power $p$ and found that the best fits to the filament profiles were obtained with $p\approx2$, so we fix this parameter to $p=2$ in the following. \citet{ArzoumanianEtAl2011} and \citet{SmithGloverKlessen2014} in observations and simulations, respectively, also found that $p=2$ gives reasonable fits, unlike the steeper profiles with $p=4$, which would represent an isothermal filament in hydrostatic equilibrium \citep{Ostriker1964}. This seems to be excluded by the observations and we also exclude such very steep profiles in all our simulations. We explain the $p=2$ slope with a simple theoretical model in which two planar shocks overlap to form a filament at their intersection line (\S\ref{sec:p}).

The Gaussian filament profile is defined as
\begin{equation}
\Sigma(r) = \Sigma(0) \exp{\left(-\frac{r^2}{2\sigma_\mathrm{Gauss}^2}\right)} + \Sigma_\mathrm{offset},
\label{eq:gauss}
\end{equation}
with the filament width $W = 2\sqrt{2\ln 2}\,\sigma_\mathrm{Gauss}\approx2.355\,\sigma_\mathrm{Gauss}$ defined as the FWHM of the Gaussian. This roughly corresponds to $W\approx3\,\rflat$ of the Plummer profile with $p=2$, Equation~(\ref{eq:plummer}), as defined in \citet{ArzoumanianEtAl2011} and confirmed here. The column density offset is a fit parameter and results in $\Sigma_\mathrm{offset}\sim3$--$5\times10^{21}\,\cm^{-2}$, close to the background column density in both the Plummer and Gaussian fits.

We note that the filament widths obtained by fitting the column density profiles depends slightly on the fit range \citep{SmithGloverKlessen2014}. It is important to choose the fit range such that the data are fitted well around the core of the filament profiles. There are two reasons for this: 1) the core of the filament profile provides the main contribution to the filament, 2) if large radii are allowed to be included in the fit, then the fit will contain contributions from overlapping filaments that are connected or close to the main filament for which the profile is computed. These overlapping contributions from multiple filaments lead to a systematically increased column density in the outer parts of the profiles \citep[see also][]{JuvelaEtAl2012b}, which artificially broadens the filament profiles and leads to an overestimate of the width. In order to avoid this problem, we constrain the fit range to $[-0.10,0.10]\,\pc$.

\begin{figure*}
\centerline{\includegraphics[width=0.63\linewidth]{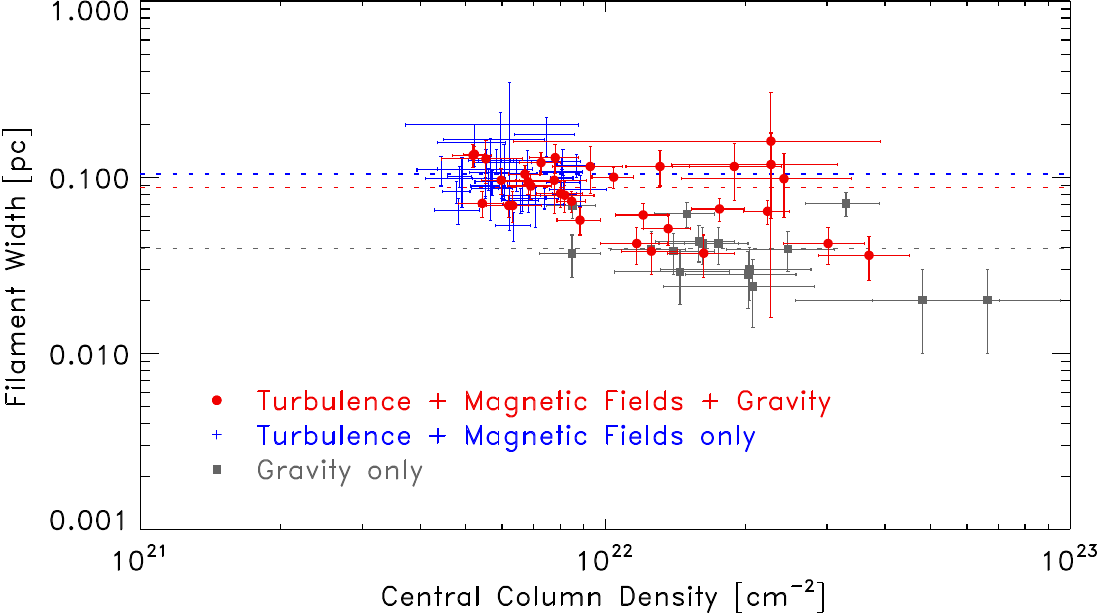}}
\caption{Filament width as a function of column density in three of our simulations: Gravity only (squares), Turbulence + Magnetic Fields only (crosses), and Turbulence + Magnetic Fields + Gravity (circles). The horizontal dotted lines show the average filament width. Gravity-only models produce narrow filaments, while turbulence models produce widths consistent with observations. Turbulence alone produces relatively low-column density filaments, while adding gravity leads to denser filaments, similar to the difference between the quiescent Polaris and the star-forming Aquila cloud shown in Figure~7 in \citet{ArzoumanianEtAl2011}.}
\label{fig:wocd}
\end{figure*}

In summary, we see that the filament profiles shown in Figure~\ref{fig:filprofs} reveal a remarkably universal filament width of about $\wfil\sim0.10\,\pc$ for models that include turbulence (panels c--f), while gravity-only models (panels a and b) have significantly narrower profiles with $\sim0.05\,\pc$. Thus, we conclude that the universality of the filament width in our simulations must be primarily the result of turbulence, with gravity merely increasing the column density of the filaments, but not significantly affecting their widths. In contrast, gravity alone is insufficient to explain the observed filament width of $\sim0.1\,\pc$.

We quantify the differences between gravity-only models and models with turbulence further, by showing the individual filament widths as a function of filament central column density in Figure~\ref{fig:wocd}. Thus, the $0.1\,\pc$ width also seen in very diffuse clouds such as Polaris can be explained by turbulence alone, while the higher-column density filaments---yet also having $0.1\,\pc$ width---for example seen in star-forming, denser clouds such as Aquila, are a result of the interplay between turbulence and gravity. This trend is clearly seen in Figure~7 in \citet{ArzoumanianEtAl2011}, where the authors compare the filament width and column density in Polaris, IC~5146 and Aquila. We find the same trend here in Figure~\ref{fig:wocd}: the filament width stays the same, but the column density is enhanced by factors of a few when the clouds are self-gravitating. The presence of a magnetic field and of feedback---while certainly affecting the number and spatial distribution of filaments (cf.~Figure~\ref{fig:images})---does not seem to change the characteristic properties of the filaments significantly (comparing panels d--f in Figure~\ref{fig:filprofs}). We provide a simple theoretical model for this universality of filament widths, which we derive from the turbulent sonic scale in Section~\ref{sec:physics}.

\subsection{Comparison with observations} \label{sec:obs}

\begin{figure*}
\centerline{\includegraphics[width=0.6\linewidth]{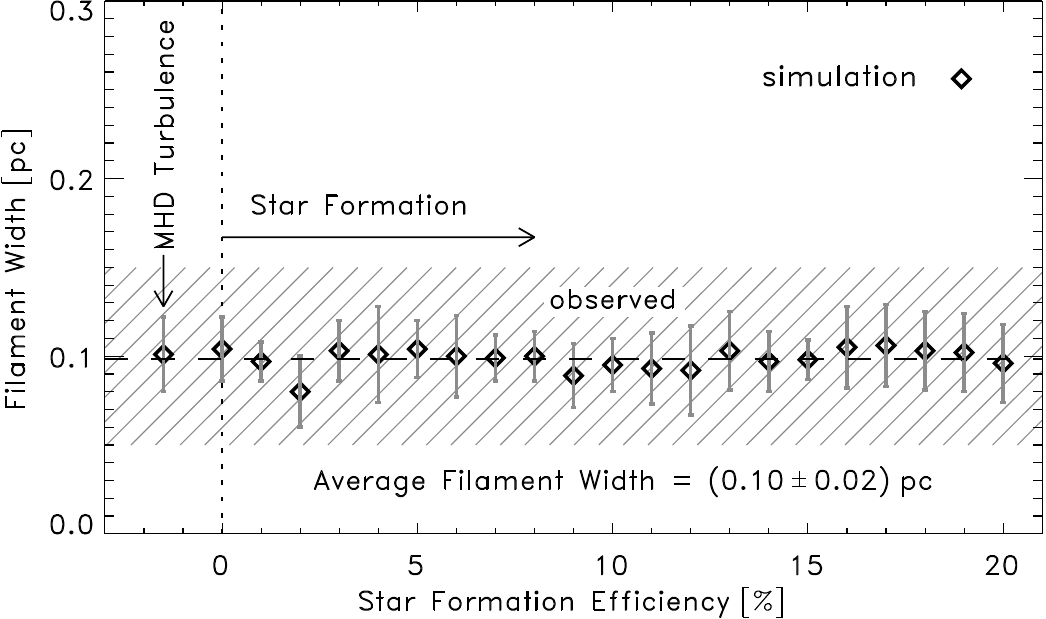}}
\caption{Filament width as a function of the star formation efficiency (SFE) in our simulation that includes gravity, turbulence, magnetic fields and jet/outflow feedback (model~6 in Table~\ref{tab:sims}), shown as diamonds. The hatched region shows the observed range of filament widths and column densities from \citet{ArzoumanianEtAl2011}, \citet{MalinenEtAl2012}, \citet{PalmeirimEtAl2013}, \citet{BenedettiniEtAl2015} and \citet{KainulainenEtAl2015}. The simulation results are in excellent agreement with the observations and do not show a significant variation with SFE. Section~\ref{sec:physics} provides a physical explanation for the universality of the filament width based on the sonic scale in MHD cloud turbulence. Note that pure MHD turbulence (labelled with an arrow at $\mathrm{SFE}<0$)---without gravity or other physical ingredients---already produces filaments with $\sim0.1\,\pc$ width.}
\label{fig:tevol}
\end{figure*}

In Figure~\ref{fig:tevol} we compare our full simulation model including gravity, turbulence, magnetic fields and jet/outflow feedback to observations. We show the filament width in the simulation as a function of the star formation efficiency (SFE), i.e., the gas fraction of the clouds that forms stars. We see that the width of the filaments is remarkably constant with $\wfil=0.10\pm0.02\,\pc$ over time and virtually independent of star formation activity. This is even more emphasised by the fact that pure MHD turbulence produces the same width, even without including gravity (indicated with an arrow for the datapoint at $\mathrm{SFE}<0$).   The simulations are in very good agreement with the currently available range of filament widths seen in observations \citep{ArzoumanianEtAl2011,JuvelaEtAl2012a,MalinenEtAl2012,PalmeirimEtAl2013,BenedettiniEtAl2015,KainulainenEtAl2015}. The agreement remains, even if we reduce the simulation resolution (see the simulation model with a maximum effective resolution of $512^3$ grid cells in Table~\ref{tab:sims} and Appendix Figure~\ref{fig:filprofs_res}).

Moreover, the agreement of our simulations with the observations is not affected by the coarser resolution of the observations. For instance, if we smooth the simulation maps to the Herschel SPIRE resolution obtained in \citet{ArzoumanianEtAl2011} as we have done in Appendix Figure~\ref{fig:images_smooth}, we still find a filament width that is statistically indistinguishable from what we obtain with the full simulation resolution.

This universality of the filament width is quite remarkable and calls for a theoretical explanation based on the physics of turbulence. We now provide a simple theoretical model for the magneto-sonic scale in order to explain the characteristic filament width of $0.1\,\pc$.

\section{The origin of the filament width, velocity and density profile} \label{sec:physics}

\subsection{Universal filament width}

The observations and the results of our turbulence simulations in Figures~\ref{fig:filprofs}--\ref{fig:tevol} show a nearly universal filament width of about $0.1\,\pc$. Here we provide a physical explanation for these observations. Both \citet{ArzoumanianEtAl2011} and \citet{FederrathDuvalKlessenSchmidtMacLow2010} suggested that the sonic scale might provide a natural, almost universal scale of filament and dense-core formation. Here we explain, derive and compute the sonic scale in detail and we compare it to observations.

The sonic scale marks the transition from supersonic to subsonic turbulence \citep{VazquezBallesterosKlessen2003}. It is the characteristic scale in a turbulent medium, such as the interstellar medium, on which the Mach number becomes unity. A fundamental property of any turbulent flow is that it exhibits a scale-dependent velocity dispersion $\sigma_v(\ell)\propto\ell^{\alpha}$, which roughly follows a power law. In the classical \citet{Kolmogorov1941c} turbulence, which strictly only applies to incompressible gases, this power law is given by $\sigma_v(\ell)\propto\ell^{1/3}$ \citep[for a review of Kolmogorov turbulence, see e.g.,][]{Frisch1995}. In contrast, the interstellar medium and especially the molecular cold phase, in which filaments and stars form, is highly compressible and supersonic, which means that the Kolmogorov theory cannot be applied to the scales where the turbulence is supersonic. Instead, \citet{Burgers1948} turbulence, which essentially consists of an ensemble of discontinuities or shocks is a much better description for the supersonic, highly compressible scales in a turbulent cloud. In this regime, numerical simulations find a steeper power-law dependence, $\sigma_v(\ell)\propto\ell^{1/2}$ \citep{KritsukEtAl2007,SchmidtEtAl2009,FederrathDuvalKlessenSchmidtMacLow2010,Federrath2013} than in Kolmogorov turbulence.

This power-law scaling of the turbulence, $\sigma_v(\ell)\propto\ell^{\alpha}$ with $\alpha>0$, implies that the turbulent velocity fluctuations $\sigma_v$ decrease with decreasing scale $\ell$. Eventually there must be a characteristic scale on which the turbulent velocity dispersion equals the sound speed, $\sigma_v(\lambda_\mathrm{sonic})=\cs$ and this implicitly defines the sonic scale $\lambda_\mathrm{sonic}$ (we provide the explicit definition below). Since the density fluctuations in a super-Alfv\'enic turbulent medium are roughly proportional to the square of the Mach number, $\mach^2=\sigma_v^2/\cs^2$, the turbulent density fluctuations will quickly vanish near and below the sonic scale. This means that the incompressible (no density fluctuations) Kolmogorov theory then provides a good approximation of the turbulence below the sonic scale, with a power-law exponent of $\alpha\sim1/3$, while scales larger than $\lambda_\mathrm{sonic}$ are controlled by supersonic turbulence with an exponent $\alpha\sim1/2$. We clearly see that the sonic scale is a fundamental characteristic scale on which the turbulence changes behaviour from being highly compressible and supersonic with $\alpha\sim1/2$ for $\ell>\lambda_\mathrm{sonic}$, to subsonic, nearly incompressible with $\alpha\sim1/3$ for $\ell<\lambda_\mathrm{sonic}$. This transition may in fact be the same as what has been termed `the transition to coherence' in observations of dense cores \citep{GoodmanEtAl1998,PinedaEtAl2010}.

Based on this, the sonic scale is explicitly defined as \citep[see Equation~(22) in][]{FederrathKlessen2012},
\begin{equation} \label{eq:ls}
\lambda_\mathrm{sonic} = L\left[\frac{\cs}{\sigma_V}\left(1+\beta^{-1}\right)^{1/2}\right]^2,
\end{equation}
where $L$, $\sigma_V$, $\cs$ and $\beta$ are the cloud scale, the velocity dispersion on the cloud scale, the sound speed, and the ratio of thermal to magnetic pressure, plasma $\beta=p_\mathrm{thermal}/p_\mathrm{magnetic}$. Note that Equation~(\ref{eq:ls}) only takes magnetic pressure into account, while magnetic tension is ignored, which would require a (so far uncertain) correction for magnetic field anisotropies. However, as long as the turbulence producing the filaments remains super- to trans-Alfv\'enic, magnetic pressure is the only significant magnetic contribution, which is covered by our expression for the sonic scale. Equation~(\ref{eq:ls}) can be evaluated at any given scale for which a velocity dispersion is available. Computing the sonic scale from the standard linewidth-size relation, $\sigma_V(L) \sim 1\,\km\,\s^{-1} (L/\pc)^{0.5}$ \citep{Larson1981,SolomonEtAl1987,OssenkopfMacLow2002,HeyerBrunt2004,HeyerEtAl2009,RomanDuvalEtAl2011}, and taking into account its variations, as well as the typical range of magnetic field strengths \citep[leading to \mbox{$\beta\sim0.3$--$\infty$}, see][]{FalgaroneEtAl2008,CrutcherEtAl2010}\footnote{$\beta\to\infty$ is the limit in which the magnetic field is zero.}, we find a relatively narrow range of sonic scales inside molecular clouds, \mbox{$\lambda_\mathrm{sonic}=0.04$--$0.16\,\pc$}.

Equation~(\ref{eq:ls}) implies that the sonic scale depends on the linewidth-size relation $\sigma_v(L)$. From this we can conclude that the observed narrow range of filament widths is a result of the relatively universal character of the linewidth-size relation in molecular clouds in the Milky Way. However, the variations of this relation then also provide a natural explanation for the range and variation of the observed filament width. Equation~(\ref{eq:ls}) further makes a direct theoretical prediction that the filament width may be systematically different in regions governed by a different linewidth-size relation. This might be the case in the centres of galaxies, such as in our Central Molecular Zone \citep{ShettyEtAl2012} or in other extreme molecular cloud conditions that alter $\sigma_v$ and $L$ to be different from the standard linewidth-size relation. Equation~(\ref{eq:ls}) covers these possibilities and it furthermore covers the dependences on the magnetic field strength (through plasma $\beta$) and on the thermal state of the gas (through the sound speed $\cs$)\footnote{Whether Equation~(\ref{eq:ls}) might also be applied to cosmic filaments such as recently studied in \citet{ButlerEtAl2015}, \citet{GhellerEtAl2015}, and \citet{TremblayEtAl2015}, remains an open question, yet certain properties of cosmic gas such as the higher temperatures would indeed result in a larger sonic scale and thus in wider filaments.}.

Moreover, Equation~(\ref{eq:ls}) may explain the somewhat increased filament widths found in the simulations by \citet{SmithGloverKlessen2014}. They use decaying turbulence simulations, which means that the Mach number drops significantly over time and especially in the initial transient phase of the simulations when the first shocks form (their initial Mach number is $\sigma_v/\cs\sim13$). Private communication with R.~Smith leads us to conclude that the Mach number has dropped to about \mbox{$\sigma_v/\cs\sim5$--$6$} when \citet{SmithGloverKlessen2014} analyse their filament profiles, which means that their simulations have evolved off the standard linewidth-size relation by that time. Given their analysis box scale $L=9.7\,\pc$, and $\beta\to\infty$ (they did not have a magnetic field), we find through Equation~(\ref{eq:ls}) that their sonic scale is around \mbox{$\lambda_\mathrm{sonic}\sim9.7\,\pc/(5$--$6)^2\sim0.27$--$0.39\,\pc$}, in reasonable agreement with their average filament widths of \mbox{$0.2$--$0.3\,\pc$}. Even though the prediction by Equation~(\ref{eq:ls}) matches the filament widths in \citet{SmithGloverKlessen2014} reasonably well, we emphasise that Equation~(\ref{eq:ls}) strictly speaking only applies to isothermal turbulence, while \citet{SmithGloverKlessen2014} include a complex heating and cooling balance in their simulations, which only produces a nearly isothermal gas over a limited range of densities with a spatially varying sound speed \citep{GloverFederrathMacLowKlessen2010}. We would expect this to produce a range of sonic scales rather than a single one, leading to a wider range of possible filament widths, which is indeed what \citet{SmithGloverKlessen2014} see in their simulations.

We can go one step further and use Equation~(\ref{eq:ls}) to explain the recent observations by \citet{WangEtAl2015} of large-scale filaments in the spiral arms of the Milky Way. Interestingly, \citet{WangEtAl2015} find that the largest filaments in the Milky Way are \mbox{$37$--$99\,\pc$} long and $0.6$--$3.0\,\pc$ wide\footnote{Note that the beam resolution of the observations by \citet{WangEtAl2015} corresponds to \mbox{$\sim\!0.4$--$0.7\,\pc$}, such that the lower limit of their range of filament widths (\mbox{$0.6$--$3.0\,\pc$}) might be affected by insufficient resolution.}, i.e., significantly wider than the $0.1\,\pc$ found for small-scale filaments inside molecular clouds (cf.~Sec.~\ref{sec:obs}). They also measure velocity dispersions of $\sigma_v=1.4$--$3.1\,\km\,\s^{-1}$ and dust temperatures of \mbox{$17$--$47\,\mathrm{K}$} for their sample of filaments. Assuming that the dust temperature is about the same as the gas temperature, we can use these measurements to compute the sound speed $\cs$ in Equation~(\ref{eq:ls}). We further assume that the velocity dispersion $\sigma_v$ measured for their filaments is dominated by the scale that corresponds to the length $L$ of the filaments\footnote{This is a reasonable assumption, because the velocity dispersion in a turbulent medium is dominated by the largest scale considered in the measurement, which is the filament length in the case of the observations by \citet{WangEtAl2015}.}. We can now insert the observational data by \citet{WangEtAl2015} into Equation~(\ref{eq:ls}) and find an average sonic scale of $\lambda_\mathrm{sonic}=0.7$--$3.5\,\pc$, which is in very good agreement with their filament widths of $0.6$--$3.0\,\pc$. This implies that there might be another characteristic sonic scale, which corresponds to the transition from the atomic to the molecular phase in the ISM. That sonic scale of about $1$--$3\,\pc$ would then be associated with molecular cloud formation, while the smaller characteristic sonic scale of $0.1\,\pc$ is associated with filament and dense-core formation inside molecular clouds.

In summary, the theoretical prediction for the range of sonic scales provided by Equation~(\ref{eq:ls}) is in very good agreement with the range of small-scale filament widths of \mbox{$0.05$--$0.15\,\pc$} found in the observations by \citet{ArzoumanianEtAl2011}, \citet{MalinenEtAl2012}, \citet{PalmeirimEtAl2013}, \citet{BenedettiniEtAl2015}, and \citet{KainulainenEtAl2015} in various different Milky Way clouds. Equation~(\ref{eq:ls}) may further explain the larger filament widths of about $0.6$--$3.0\,\pc$ seen in the observations by \citet{WangEtAl2015}. The smaller characteristic sonic scale of about $0.1\,\pc$ represents the typical scale of dense-core formation inside molecular clouds, while the larger sonic scale of about $1$--$3\,\pc$ might correspond to molecular cloud formation from the atomic phase, i.e., a characteristic scale for the atomic-to-molecular transition in the ISM. While we cannot rule out other possible explanations, the sonic scale provides an encouraging simple and plausible explanation for the observed characteristic widths of interstellar filaments.

\subsection{Filament velocity dispersion}

\begin{figure*}
\centerline{\includegraphics[width=0.6\linewidth]{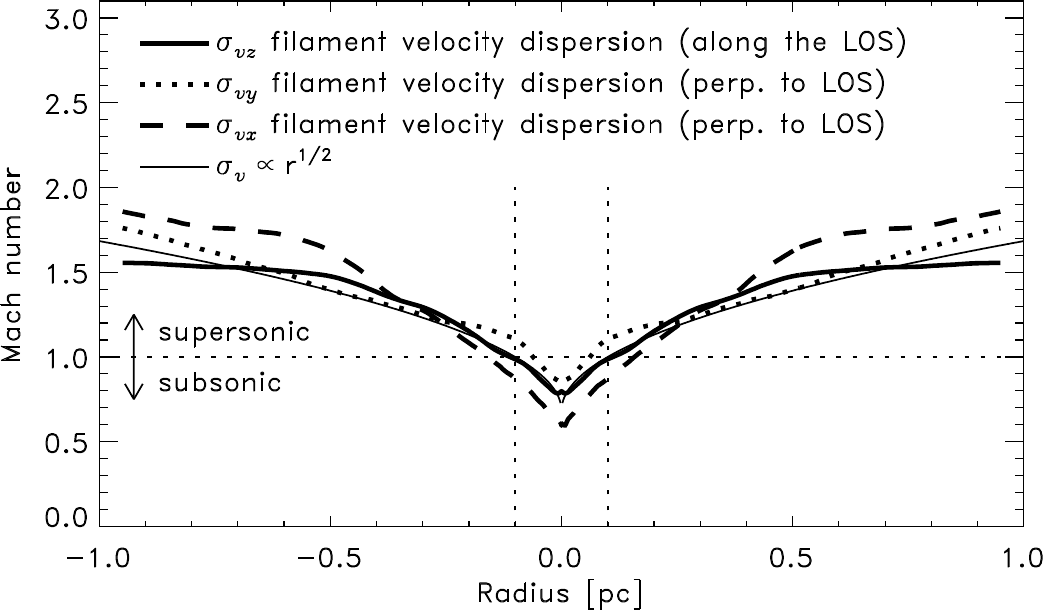}}
\caption{Filament velocity dispersion profiles along the line-of-sight (LOS), $\sigma_{vz}$ (solid line), and perpendicular to the LOS, $\sigma_{vy}$ (dotted line), and $\sigma_{vx}$ (dashed line) in our simulation that includes gravity, turbulence, magnetic fields and jet/outflow feedback (model~6 in Table~\ref{tab:sims}). The velocity dispersion was normalised to the sound speed, such that these curves show the Mach number inside and outside the filaments as a function of radius centred on the filaments. The thin solid line shows the typical scaling for molecular cloud turbulence, $\sigma_v\propto r^{1/2}$ \citep[see e.g.,][and references therein]{Federrath2013}. We see that the velocity profiles roughly follow this scaling. Most importantly, the turbulence is trans- to subsonic inside the filament and supersonic outside (with the respective regions separated by thin dotted lines), which is the key assumption behind the idea that the filament width is determined by the sonic scale, Equation~(\ref{eq:ls}).}
\label{fig:vprof}
\end{figure*}

We see that our theoretical model for the filament width based on the sonic scale, Equation~(\ref{eq:ls}), implies that the filament velocity dispersion should be trans- to subsonic inside the filaments and supersonic outside. In order to test this, we show the velocity dispersion profiles of our filaments in the simulation that includes gravity, turbulence, magnetic fields and jet/outflow feedback (model~6 in Table~\ref{tab:sims}) in Figure~\ref{fig:vprof}. Indeed, we find that the Mach number (the ratio of velocity dispersion to sound speed) is below unity inside the filaments and greater than unity outside, consistent with what is found in a recent observation of the Musca filament \citep{HacarEtAl2015}. We further see that the velocity dispersion outside the filaments roughly follows the typical scaling of supersonic turbulence, $\sigma_v\propto r^{1/2}$, which is the essential ingredient in the derivation of Equation~(\ref{eq:ls}) for the sonic scale.

\subsection{The origin of the $\Sigma\propto r^{-p+1}$ and $\rho\propto r^{-p}$ filament profile scaling with $p=2$} \label{sec:p}

Figure~\ref{fig:filprofs} showed that the filament column density profiles follow a scaling of $\Sigma\propto r^{-1}$, i.e., they are best fit with a Plummer-profile exponent of $p=2$ in Equation~(\ref{eq:plummer}). Note that this implies that the filament volume density scales as $\rho\propto\Sigma/r\propto r^{-p}$, i.e., $\rho\propto r^{-2}$ for $p=2$.  \citet{Ostriker1964} analysed the scaling of hydrostatic, isothermal cylinders (filaments) and found $p=4$ by assuming that the gravitational acceleration is balanced by the gas pressure gradient. In contrast, here we find $p=2$, consistent with other recent works \citep{FiegePudritz2000,ArzoumanianEtAl2011,ContrerasRathborneGaray2013,GomezVazquez2014,SmithGloverKlessen2014}. Most striking is the fact that this scaling is present in turbulence-only models (cf.~panel~c in Fig.~\ref{fig:filprofs}). In other words, turbulence alone must be sufficient to yield such a scaling of $\rho\propto r^{-2}$ or equivalently $\Sigma\propto r^{-1}$. Here we provide a simple physical model based on turbulent compression in shocks to explain this nearly universal $p=2$ scaling of the filament density profile.

In any planar (2D) radiative shock (purely hydrodynamic or MHD), the post-shock density $\rho$ scales inversely with the post-shock thickness $\lambda$, as $\rho\propto\lambda^{-1}$. A filament occurs where two planar shocks intersect, i.e., the intersection of two planes is a line (filament). Thus, at the location of the filament, two planar shocks (index 1 and 2) collide or intersect, such that the density of the filament scales as
\begin{equation} \label{eq:rhoscaling}
\rho_\mathrm{filament}\propto\rho_\mathrm{shock,1}\rho_\mathrm{shock,2}\propto\lambda_1^{-1}\lambda_2^{-1}\sim\lambda^{-2},
\end{equation}
because shock collision is a multiplicative process in the density \citep[e.g.,][]{Vazquez1994}. Equation~(\ref{eq:rhoscaling}) thus provides a simple geometric argument for why the filament density profile scales with $r^{-2}$, i.e., $p=2$, based on the collision of two shocks forming a filament. We see that this does not require gravity. It is solely the result of hydrodynamic (or MHD) interactions of shocks and these shocks are the hallmark of supersonic turbulence in molecular clouds.

\subsection{Discussion of previous filament models}

As explained in the preceding section, \citet{Ostriker1964} provides a model for the density profile of isothermal and non-isothermal filaments, assuming hydrostatic balance. The prediction for isothermal filaments ($p=4$) neither matches the observations nor the simulations, which consistently show that $p=2$ yields the best fit to the filament density and column density profiles. The \citet{Ostriker1964} model requires gravity and assumes hydrostatic equilibrium. Both assumptions are problematic. First, it is hard to see that structures in molecular clouds governed by supersonic turbulence should be in hydrostatic equilibrium. Instead, the filaments likely form by dynamic turbulent shock interactions. Second, quiescent clouds such as Polaris have filaments with very similar characteristics as filaments in self-gravitating, star-forming clouds. Thus, the filaments in very diffuse clouds such as Polaris, where gravity does not play a significant role, cannot be explained with the \citet{Ostriker1964} model.

\citet{FischeraMartin2012a,FischeraMartin2012b} extend the analysis by \citet{Ostriker1964} by including an external pressure, but otherwise, they use very similar assumptions, including hydrostatic equilibrium, gravity and pressure balance to derive the density profile of filaments. While this model can explain some of the properties of self-gravitating filaments, it cannot explain non-self-gravitating ones and still assumes hydrostatic equilibrium.

\citet{Heitsch2013a,Heitsch2013b} follow the same approach as in \citet{FischeraMartin2012a,FischeraMartin2012b}, but include the effects of magnetic fields.

\citet{HennebelleAndre2013} provide a theoretical model assuming self-gravitating accreting filaments. In this model, gravity is not only balanced by thermal pressure, but also by turbulent pressure and dissipation induced by ambipolar diffusion. As in the \citet{Ostriker1964}, \citet{FischeraMartin2012a,FischeraMartin2012b}, and \citet{Heitsch2013a,Heitsch2013b} models, the model by \citet{HennebelleAndre2013} cannot explain filaments in diffuse clouds primarily governed by supersonic turbulence, because it requires the filaments to be self-gravitating. This is an important limitation, because observations by \citet{ArzoumanianEtAl2011} and \citet{PanopoulouEtAl2014}, as well as our turbulence-only simulation show that filaments with the typical observed properties can already arise under conditions where self-gravity is negligible.

\section{Summary and conclusions} \label{sec:conclusions}

We compared the properties of filaments formed in a set of six high-resolution simulations following the evolution of molecular clouds and star formation within them, combining different physical ingredients: gravity, turbulence, magnetic fields, and jet/outflow feedback from young stars. Here we list our main findings and conclusions:

\begin{enumerate}

\item Using DisPerSE, we detect complex networks of filamentary structures in all our simulations  (cf.~Fig.~\ref{fig:images}).

\item We find that the radial filament profiles of the simulations that include turbulence reveal a remarkably universal filament width of about $0.10\pm0.02\,\pc$, while gravity-only simulations produce significantly narrower filaments with $\sim0.05\,\pc$ (cf.~Fig.~\ref{fig:filprofs}).

\item Pure MHD turbulence can account for the filament properties in diffuse clouds such as Polaris, while star-forming clouds such as Aquila have higher column densities with a wider distribution, yet their widths are still $\sim0.1\,\pc$. This trend is reproduced in our simulations that compare pure MHD turbulence on the one hand and MHD turbulence including gravity and star formation on the other hand (cf.~Fig.~\ref{fig:wocd}).

\item We show that the filament width in the simulations with turbulence is in excellent agreement with observations. The filament width does not systematically depend on the evolutionary stage or the star formation efficiency of the clouds (cf.~Fig.~\ref{fig:tevol}).

\item We explain the nearly universal width of interstellar filaments of $\sim0.1\,\pc$ with a theoretical model based on the scaling of supersonic, magnetised turbulence (Sec.~\ref{sec:physics}). In this model, the filament width coincides with the sonic scale, which marks the transition from the large supersonic scales towards the small subsonic scales of a molecular cloud. Equation~(\ref{eq:ls}) provides the sonic scale as a function of the cloud scale $L$, the velocity dispersion $\sigma_v$, the gas sound speed $\cs$ and the strength of the magnetic field parametrized by plasma $\beta$. Given typical molecular cloud conditions in the Milky Way, Equation~(\ref{eq:ls}) yields a sonic scale of \mbox{$\lambda_\mathrm{sonic}=0.04$--$0.16\,\pc$}, in very good agreement with the observed filament widths of \mbox{$0.05$--$0.15\,\pc$}.

\item We find that the filament velocity dispersion is trans- to subsonic inside the filaments and supersonic outside, and follows the scaling for supersonic turbulence, $\sigma_v\propto r^{1/2}$ (cf.~Fig.~\ref{fig:vprof}). This confirms the main assumptions behind the theoretical model for the filament width determined by the sonic scale, Equation~(\ref{eq:ls}).

\item We explain the $p=2$ scaling of the filament column density with radius, $\Sigma\propto r^{-p+1}$, implying a volume density scaling of $\rho\propto r^{-p}$, purely with the scaling of the post-shock density with the post-shock thickness in two colliding planar shocks, forming a filament at their intersection line (cf.~\S\ref{sec:p}).

\item Our filament widths are converged with numerical resolution (cf.~Fig.~\ref{fig:filprofs_res}) and do not depend on the telescope beam smoothing as long as the observational resolution is comparable or better than the filament width (cf.~Fig.~\ref{fig:images_smooth}). Reducing the DisPerSE persistence threshold by a factor of ten yields more filaments with lower average column densities, but the width is not significantly affected by the persistence threshold (cf.~Fig.~\ref{fig:filprofs_persistence}).

\item Finally, we find that the magnetic field does not have a preferred orientation with respect to the filaments and that the magnetic field component parallel to the filament axis is slightly enhanced inside the filament, which is caused by turbulent compression of the field during the formation of the filaments (cf.~Fig.~\ref{fig:magprof}).

\end{enumerate}

\section*{Acknowledgements}
I thank D.~Arzoumanian for sending the filament profile of IC~5146 shown in Figure~\ref{fig:ic5146comp} for comparison with our simulations and R.~Smith for providing the Mach numbers of their decaying turbulence simulations. I further thank P.~Andr\'e, D.~Arzoumanian, R.~Crocker, M.~Cunningham, C.~Green, E.~Hansen, E.~Kaminsky, N.~Schneider, R.~Smith and E.~Vazquez-Semadeni for interesting discussions on filaments and comments on the manuscript, as well as the anonymous referee for their critical comments, which improved this work.
C.F.~acknowledges funding provided by the Australian Research Council's Discovery Projects (grants~DP130102078 and~DP150104329).
I gratefully acknowledge the J\"ulich Supercomputing Centre (grant hhd20), the Leibniz Rechenzentrum and the Gauss Centre for Supercomputing (grants~pr32lo, pr48pi and GCS Large-scale project~10391), the Partnership for Advanced Computing in Europe (PRACE grant pr89mu), the Australian National Computational Infrastructure (grant~ek9), and the Pawsey Supercomputing Centre with funding from the Australian Government and the Government of Western Australia.
The software used in this work was in part developed by the DOE-supported Flash Center for Computational Science at the University of Chicago.

\appendix

\section{Resolution study} \label{app:res}

Here we provide a resolution study, both for the simulations as well as for the synthetic column density maps.

\subsection{Simulation resolution} \label{app:resnum}

\begin{figure}
\centerline{\includegraphics[width=1.0\linewidth]{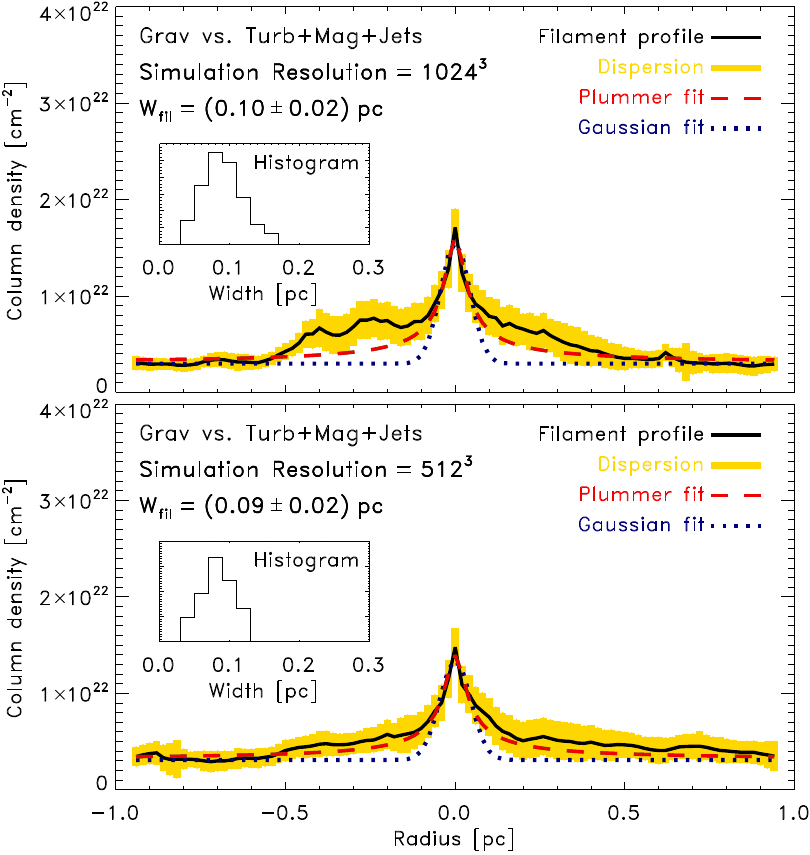}}
\caption{Same as Figure~\ref{fig:filprofs}, but for different simulation resolutions: our standard resolution of $1024^3$ grid cells (top panel) vs.~$512^3$ cells (bottom panel). We see that the filament width does not depend systematically on the numerical resolution, thus demonstrating convergence of the filament width.}
\label{fig:filprofs_res}
\end{figure}

Figure~\ref{fig:filprofs_res} shows the filament profiles in simulation `Grav vs.~Turb+Mag+Jets' for two different maximum grid resolutions: $1024^3$ (top panel) vs.~$512^3$ (bottom panel). We find that the filament profiles do not depend significantly on the numerical resolution of the simulations. The measured filament widths (both from the stacked profile and from individual filament profiles) only vary within the fit uncertainties.

\subsection{Observational resolution} \label{app:resobs}

\begin{figure*}
\centerline{\includegraphics[width=1.0\linewidth]{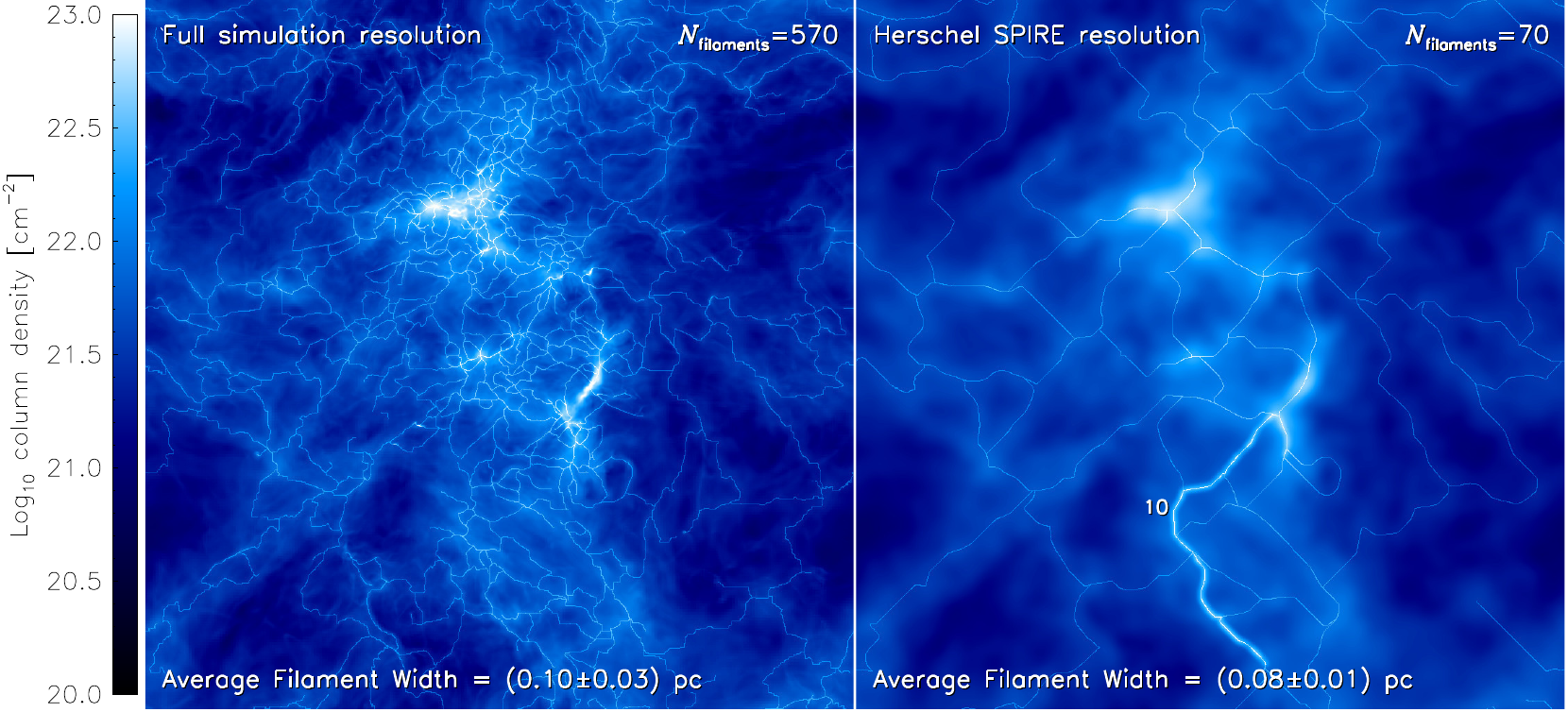}}
\caption{Comparison of synthetic column density maps in our full simulation resolution (left-hand panel) and smoothed to the Herschel SPIRE ($250\,\mu\mathrm{m}$) resolution of $0.04\,\pc$ at the distance of $\sim\!460\,\pc$ to IC~5146 obtained in \citet{ArzoumanianEtAl2011}. To facilitate the comparison, the persistence threshold was set to $5\times10^{20}\,\cm^{-2}$, the same as in \citet{ArzoumanianEtAl2011}. While the number of detected filaments clearly depends on the telescope beam resolution, the resolution in \citet{ArzoumanianEtAl2011} is sufficient to obtain converged filament widths. In the right-hand panel, we highlight one particular filament (labelled filament~10), for which we provide a direct comparison of the filament profile with filament~6 in IC~5146 from \citet{ArzoumanianEtAl2011} (see Figure~\ref{fig:ic5146comp}).}
\label{fig:images_smooth}
\end{figure*}

In order to investigate the effects of a finite telescope resolution, we apply a Gaussian beam smoothing to our synthetic column density map of simulation `Grav vs.~Turb+Mag+Jets' in Figure~\ref{fig:images_smooth}. The left-hand image shows the column density map with filaments highlighted in our full simulation resolution and the right-hand panel shows the same map smoothed to the Herschel SPIRE ($250\,\mu\mathrm{m}$) resolution ($0.04\,\pc$) as in the observations by \citet{ArzoumanianEtAl2011}. To facilitate the comparison, we here set the persistence threshold to $5\times10^{20}\,\cm^{-2}$, the same as in \citet{ArzoumanianEtAl2011}.

Beam smoothing clearly results in the detection of less filaments, because of the loss of small-scale structures. The filament width of those filaments that are detected, however, yield a similar filament width of about $0.08\pm0.01\,\pc$ in the smoothed map as in the full-resolution map ($0.10\pm0.03\,\pc$). Thus, we conclude that the finite resolution of the observations in \citet{ArzoumanianEtAl2011} did not affect their measured filament widths. What is clearly affected though, is the total number of detected filaments and their average length, because these strongly depend on the beam resolution of the observation. We thus expect to see significantly more substructure with many more filaments in higher-resolution observations to come in the future.

\begin{figure*}
\centerline{\includegraphics[width=1.0\linewidth]{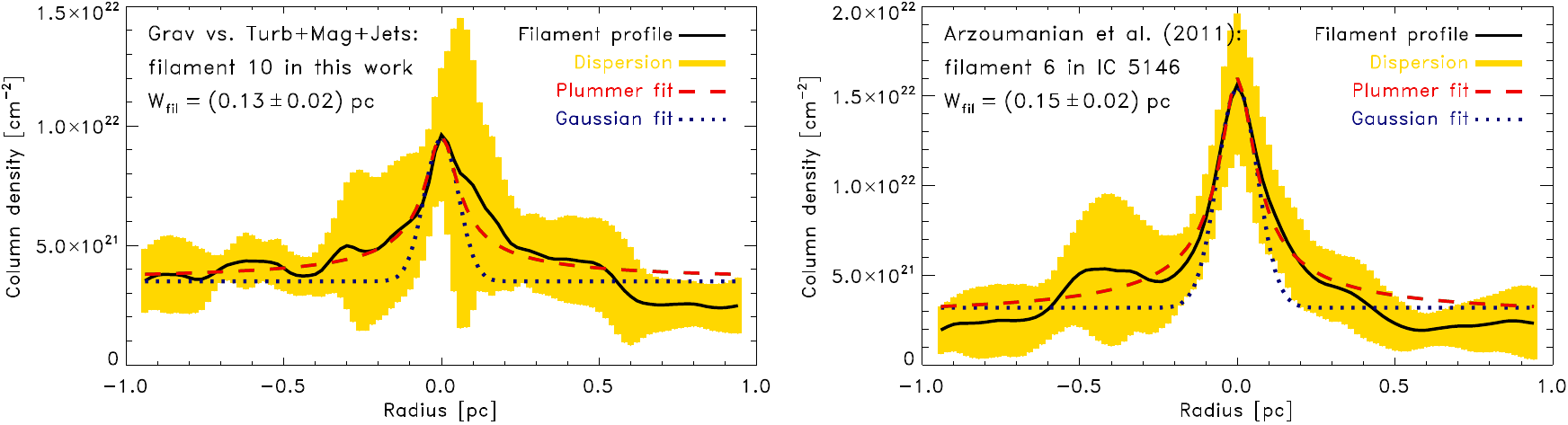}}
\caption{Comparison of the filament profile for simulation filament~10 (left-hand panel) highlighted in Figure~\ref{fig:images_smooth} and observational filament~6 in IC~5146 from \citet{ArzoumanianEtAl2011} (right-hand panel). We find similar features in the profiles from both simulations and observations, with comparable widths, column densities and profile structures containing multiple overlapping side-filaments.}
\label{fig:ic5146comp}
\end{figure*}

In order to get a feeling for the similarities and differences in the individual filament profiles between the simulations and the observations, we show in Figure~\ref{fig:ic5146comp} the profile of filament~10 detected in our simulation map from Figure~\ref{fig:images_smooth} side-by-side with filament~6 in IC~5146 from \citet{ArzoumanianEtAl2011}. The structure of both simulation and observational filament profiles is comparable, with similar widths, column densities and various overlapping side-filaments producing column-density excesses (bumps) on both sides of the filament main peaks. These overlapping nearby filaments contribute to the main filament profile and systematically increase the column density on both sides of the profile maximum. This shows why it is important to constrain the Plummer and Gaussian fits to a relatively narrow range around the filament core, in order to avoid overestimating the filament width by accidentally including overlapping contributions from other nearby filaments (see the detailed discussion in Section~\ref{sec:profiles}).

\section{Influence of the persistence threshold} \label{app:pers}

\begin{figure}
\centerline{\includegraphics[width=1.0\linewidth]{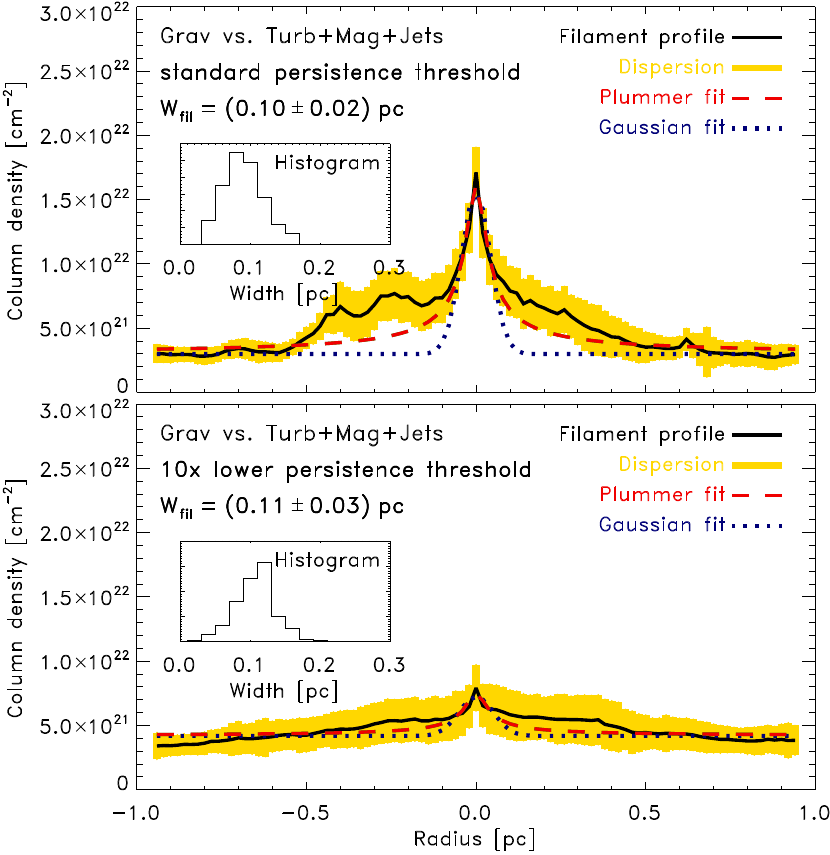}}
\caption{Influence of the DisPerSE persistence threshold for the detection of filaments (see Section~\ref{sec:disperse} for a summary of the DisPerSE algorithm). The top panel shows the filament profile based on our standard persistence threshold, the mean column density. The bottom panel shows the same, but for a ten times lower persistence threshold, as in \citet{ArzoumanianEtAl2011}. While the average filament column density clearly moves down with decreasing persistence threshold, we find that the average width does not systematically depend on the persistence threshold.}
\label{fig:filprofs_persistence}
\end{figure}

The persistence threshold in the DisPerSE algorithm for filament detection is the only free parameter, but it is also the key parameter in the algorithm. The persistence threshold basically controls which structures are taken into account when finding filaments, which means that the resulting number of detected filaments naturally depends on the choice of the persistence threshold. \citet{ArzoumanianEtAl2011} chose a persistence threshold of $5\times10^{20}\,\cm^{-2}$, primarily determined by their signal-to-noise level. Now the question is whether the filament width depends on this particular choice or not.

In Figure~\ref{fig:filprofs_persistence} we show filament profiles obtained from simulation maps with our standard persistence threshold, which is the mean column density (top panel), compared to the one chosen by \citet{ArzoumanianEtAl2011}, which is ten times lower (bottom panel). Our comparison shows that the filament peak column density systematically increases with the persistence threshold, but that the filament width is a robust measure, because it does not depend significantly on the choice of the persistence threshold.

\section{Filament magnetic field profiles and orientation} \label{sec:mags}

\begin{figure}
\centerline{\includegraphics[width=1.0\linewidth]{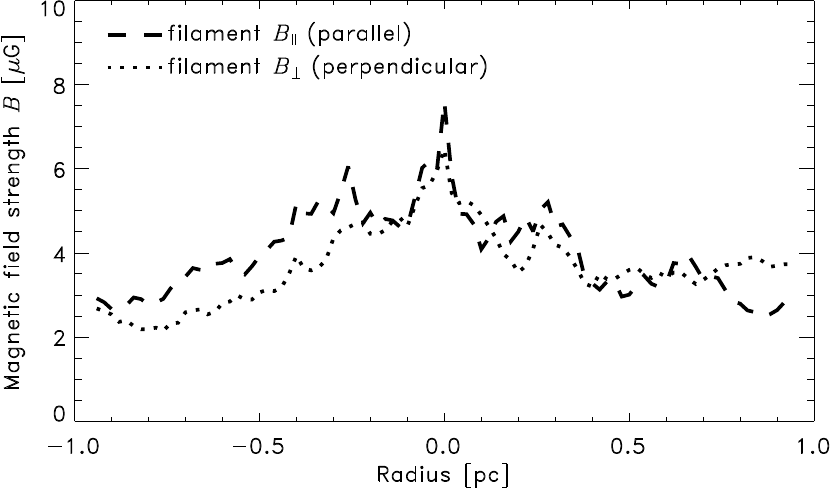}}
\caption{Filament magnetic field profiles parallel ($B_\parallel$) and perpendicular ($B_\perp$) to the filament axis, in simulation model~6 (see Tab.~\ref{tab:sims}). We find that there is no preferred orientation of the magnetic field with respect to the filaments. We also see that $B_\parallel$ is somewhat enhanced inside the filaments, which is caused by the turbulent compression of the magnetic field during the formation of the filaments.}
\label{fig:magprof}
\end{figure}

Figure~\ref{fig:magprof} shows the average magnetic field profile of the filaments in the most realistic simulation, i.e., the one that includes gravity, turbulence, magnetic fields and jet/outflow feedback (model~6 in Table~\ref{tab:sims}). We distinguish the magnetic field strength parallel ($B_\parallel$) and perpendicular ($B_\perp$) to the filament in the plane of the sky, i.e., what would be obtained from comparing the polarisation angle with the filament orientation in an observation. Figure~\ref{fig:magprof} shows that $B_\parallel$ and $B_\perp$ are similar, which means that the magnetic field does not have a significant preferred systematic orientation with respect to the filament. This indicates that the magnetic field does not play a dominant role in the filament formation process, consistent with the fact that we obtain similar filament properties in the simulation without magnetic field (panel d in Figure~\ref{fig:filprofs}). We also find in Figure~\ref{fig:magprof} that the magnetic field component parallel to the filament axis is somewhat enhanced inside the filaments, due to compression of the magnetic field lines inside the filaments.


\begin{thebibliography}{105}
\expandafter\ifx\csname natexlab\endcsname\relax\def\natexlab#1{#1}\fi

\bibitem[{{Andr{\'e}} {et~al.}(2014){Andr{\'e}}, {Di Francesco},
  {Ward-Thompson}, {Inutsuka}, {Pudritz}, \& {Pineda}}]{AndreEtAl2014}
{Andr{\'e}}, P., {Di Francesco}, J., {Ward-Thompson}, D., {et~al.} 2014,
  Protostars and Planets VI, 27

\bibitem[{{Andr{\'e}} {et~al.}(2010){Andr{\'e}}, {Men'shchikov}, {Bontemps},
  {K{\"o}nyves}, {Motte}, {Schneider}, {Didelon}, {Minier}, {Saraceno},
  {Ward-Thompson}, {di Francesco}, {White}, {Molinari}, {Testi}, {Abergel},
  {Griffin}, {Henning}, {Royer}, {Mer{\'{\i}}n}, {Vavrek}, {Attard},
  {Arzoumanian}, {Wilson}, {Ade}, {Aussel}, {Baluteau}, {Benedettini},
  {Bernard}, {Blommaert}, {Cambr{\'e}sy}, {Cox}, {di Giorgio}, {Hargrave},
  {Hennemann}, {Huang}, {Kirk}, {Krause}, {Launhardt}, {Leeks}, {Le Pennec},
  {Li}, {Martin}, {Maury}, {Olofsson}, {Omont}, {Peretto}, {Pezzuto}, {Prusti},
  {Roussel}, {Russeil}, {Sauvage}, {Sibthorpe}, {Sicilia-Aguilar}, {Spinoglio},
  {Waelkens}, {Woodcraft}, \& {Zavagno}}]{AndreEtAl2010}
{Andr{\'e}}, P., {Men'shchikov}, A., {Bontemps}, S., {et~al.} 2010, \aap, 518,
  L102

\bibitem[{{Arzoumanian} {et~al.}(2011){Arzoumanian}, {Andr{\'e}}, {Didelon},
  {K{\"o}nyves}, {Schneider}, {Men'shchikov}, {Sousbie}, {Zavagno}, {Bontemps},
  {di Francesco}, {Griffin}, {Hennemann}, {Hill}, {Kirk}, {Martin}, {Minier},
  {Molinari}, {Motte}, {Peretto}, {Pezzuto}, {Spinoglio}, {Ward-Thompson},
  {White}, \& {Wilson}}]{ArzoumanianEtAl2011}
{Arzoumanian}, D., {Andr{\'e}}, P., {Didelon}, P., {et~al.} 2011, \aap, 529, L6

\bibitem[{{Balsara} {et~al.}(2001){Balsara}, {Ward-Thompson}, \&
  {Crutcher}}]{BalsaraEtAl2001}
{Balsara}, D., {Ward-Thompson}, D., \& {Crutcher}, R.~M. 2001, \mnras, 327, 715

\bibitem[{{Benedettini} {et~al.}(2015){Benedettini}, {Schisano}, {Pezzuto},
  {Elia}, {Andr{\'e}}, {K{\"o}nyves}, {Schneider}, {Tremblin}, {Arzoumanian},
  {di Giorgio}, {Di Francesco}, {Hill}, {Molinari}, {Motte}, {Nguyen-Luong},
  {Palmeirim}, {Rivera-Ingraham}, {Roy}, {Rygl}, {Spinoglio}, {Ward-Thompson},
  \& {White}}]{BenedettiniEtAl2015}
{Benedettini}, M., {Schisano}, E., {Pezzuto}, S., {et~al.} 2015, \mnras, 453,
  2036

\bibitem[{{Berger} \& {Colella}(1989)}]{BergerColella1989}
{Berger}, M.~J., \& {Colella}, P. 1989, Journal of Computational Physics, 82,
  64

\bibitem[{{Brunt} {et~al.}(2009){Brunt}, {Heyer}, \& {Mac
  Low}}]{BruntHeyerMacLow2009}
{Brunt}, C.~M., {Heyer}, M.~H., \& {Mac Low}, M. 2009, \aap, 504, 883

\bibitem[{{Burgers}(1948)}]{Burgers1948}
{Burgers}, J.~M. 1948, Advances in Applied Mechanics, 1, 171

\bibitem[{{Butler} {et~al.}(2015){Butler}, {Tan}, \& {Van
  Loo}}]{ButlerEtAl2015}
{Butler}, M.~J., {Tan}, J.~C., \& {Van Loo}, S. 2015, \apj, 805, 1

\bibitem[{{Contreras} {et~al.}(2013){Contreras}, {Rathborne}, \&
  {Garay}}]{ContrerasRathborneGaray2013}
{Contreras}, Y., {Rathborne}, J., \& {Garay}, G. 2013, \mnras, 433, 251

\bibitem[{{Crutcher} {et~al.}(2010){Crutcher}, {Wandelt}, {Heiles},
  {Falgarone}, \& {Troland}}]{CrutcherEtAl2010}
{Crutcher}, R.~M., {Wandelt}, B., {Heiles}, C., {Falgarone}, E., \& {Troland},
  T.~H. 2010, \apj, 725, 466

\bibitem[{{Cunningham} {et~al.}(2011){Cunningham}, {Klein}, {Krumholz}, \&
  {McKee}}]{CunninghamEtAl2011}
{Cunningham}, A.~J., {Klein}, R.~I., {Krumholz}, M.~R., \& {McKee}, C.~F. 2011,
  \apj, 740, 107

\bibitem[{{Dubey} {et~al.}(2008){Dubey}, {Fisher}, {Graziani}, {Jordan},
  {Lamb}, {Reid}, {Rich}, {Sheeler}, {Townsley}, \& {Weide}}]{DubeyEtAl2008}
{Dubey}, A., {Fisher}, R., {Graziani}, C., {et~al.} 2008, in Astronomical
  Society of the Pacific Conference Series, Vol. 385, Numerical Modeling of
  Space Plasma Flows, ed. N.~V. {Pogorelov}, E.~{Audit}, \& G.~P. {Zank}, 145

\bibitem[{{Elmegreen} \& {Scalo}(2004)}]{ElmegreenScalo2004}
{Elmegreen}, B.~G., \& {Scalo}, J. 2004, \araa, 42, 211

\bibitem[{{Eswaran} \& {Pope}(1988)}]{EswaranPope1988}
{Eswaran}, V., \& {Pope}, S.~B. 1988, CF, 16, 257

\bibitem[{{Falgarone} {et~al.}(1992){Falgarone}, {Puget}, \&
  {Perault}}]{FalgaronePugetPerault1992}
{Falgarone}, E., {Puget}, J.-L., \& {Perault}, M. 1992, \aap, 257, 715

\bibitem[{{Falgarone} {et~al.}(2008){Falgarone}, {Troland}, {Crutcher}, \&
  {Paubert}}]{FalgaroneEtAl2008}
{Falgarone}, E., {Troland}, T.~H., {Crutcher}, R.~M., \& {Paubert}, G. 2008,
  \aap, 487, 247

\bibitem[{{Federrath}(2013)}]{Federrath2013}
{Federrath}, C. 2013, \mnras, 436, 1245

\bibitem[{{Federrath}(2015)}]{Federrath2015}
---. 2015, \mnras, 450, 4035

\bibitem[{{Federrath} {et~al.}(2010{\natexlab{a}}){Federrath}, {Banerjee},
  {Clark}, \& {Klessen}}]{FederrathBanerjeeClarkKlessen2010}
{Federrath}, C., {Banerjee}, R., {Clark}, P.~C., \& {Klessen}, R.~S.
  2010{\natexlab{a}}, \apj, 713, 269

\bibitem[{{Federrath} \& {Klessen}(2012)}]{FederrathKlessen2012}
{Federrath}, C., \& {Klessen}, R.~S. 2012, \apj, 761, 156

\bibitem[{{Federrath} {et~al.}(2010{\natexlab{b}}){Federrath}, {Roman-Duval},
  {Klessen}, {Schmidt}, \& {Mac Low}}]{FederrathDuvalKlessenSchmidtMacLow2010}
{Federrath}, C., {Roman-Duval}, J., {Klessen}, R.~S., {Schmidt}, W., \& {Mac
  Low}, M. 2010{\natexlab{b}}, \aap, 512, A81

\bibitem[{{Federrath} {et~al.}(2014){Federrath}, {Schr{\"o}n}, {Banerjee}, \&
  {Klessen}}]{FederrathEtAl2014}
{Federrath}, C., {Schr{\"o}n}, M., {Banerjee}, R., \& {Klessen}, R.~S. 2014,
  \apj, 790, 128

\bibitem[{{Federrath} {et~al.}(2011){Federrath}, {Sur}, {Schleicher},
  {Banerjee}, \& {Klessen}}]{FederrathSurSchleicherBanerjeeKlessen2011}
{Federrath}, C., {Sur}, S., {Schleicher}, D.~R.~G., {Banerjee}, R., \&
  {Klessen}, R.~S. 2011, \apj, 731, 62

\bibitem[{{Fiege} \& {Pudritz}(2000)}]{FiegePudritz2000}
{Fiege}, J.~D., \& {Pudritz}, R.~E. 2000, \mnras, 311, 85

\bibitem[{{Fischera} \& {Martin}(2012{\natexlab{a}})}]{FischeraMartin2012b}
{Fischera}, J., \& {Martin}, P.~G. 2012{\natexlab{a}}, \aap, 547, A86

\bibitem[{{Fischera} \& {Martin}(2012{\natexlab{b}})}]{FischeraMartin2012a}
---. 2012{\natexlab{b}}, \aap, 542, A77

\bibitem[{Frisch(1995)}]{Frisch1995}
Frisch, U. 1995, Turbulence, the legacy of A.~N.~Kolmogorov ({Cambridge Univ.
  Press})

\bibitem[{{Fryxell} {et~al.}(2000){Fryxell}, {Olson}, {Ricker}, {Timmes},
  {Zingale}, {Lamb}, {MacNeice}, {Rosner}, {Truran}, \&
  {Tufo}}]{FryxellEtAl2000}
{Fryxell}, B., {Olson}, K., {Ricker}, P., {et~al.} 2000, \apjs, 131, 273

\bibitem[{{Gaensler} {et~al.}(2011){Gaensler}, {Haverkorn}, {Burkhart},
  {Newton-McGee}, {Ekers}, {Lazarian}, {McClure-Griffiths}, {Robishaw},
  {Dickey}, \& {Green}}]{GaenslerEtAl2011}
{Gaensler}, B.~M., {Haverkorn}, M., {Burkhart}, B., {et~al.} 2011, \nat, 478,
  214

\bibitem[{{Gheller} {et~al.}(2015){Gheller}, {Vazza}, {Favre}, \&
  {Br{\"u}ggen}}]{GhellerEtAl2015}
{Gheller}, C., {Vazza}, F., {Favre}, J., \& {Br{\"u}ggen}, M. 2015, \mnras,
  453, 1164

\bibitem[{{Ginsburg} {et~al.}(2013){Ginsburg}, {Federrath}, \&
  {Darling}}]{GinsburgFederrathDarling2013}
{Ginsburg}, A., {Federrath}, C., \& {Darling}, J. 2013, \apj, 779, 50

\bibitem[{{Glover} {et~al.}(2010){Glover}, {Federrath}, {Mac Low}, \&
  {Klessen}}]{GloverFederrathMacLowKlessen2010}
{Glover}, S.~C.~O., {Federrath}, C., {Mac Low}, M., \& {Klessen}, R.~S. 2010,
  \mnras, 404, 2

\bibitem[{{G{\'o}mez} \& {V{\'a}zquez-Semadeni}(2014)}]{GomezVazquez2014}
{G{\'o}mez}, G.~C., \& {V{\'a}zquez-Semadeni}, E. 2014, \apj, 791, 124

\bibitem[{{Goodman} {et~al.}(1998){Goodman}, {Barranco}, {Wilner}, \&
  {Heyer}}]{GoodmanEtAl1998}
{Goodman}, A.~A., {Barranco}, J.~A., {Wilner}, D.~J., \& {Heyer}, M.~H. 1998,
  \apj, 504, 223

\bibitem[{{Hacar} {et~al.}(2015){Hacar}, {Kainulainen}, {Tafalla}, {Beuther},
  \& {Alves}}]{HacarEtAl2015}
{Hacar}, A., {Kainulainen}, J., {Tafalla}, M., {Beuther}, H., \& {Alves}, J.
  2015, \aap, accepted (arXiv:1511.06370)

\bibitem[{{Hacar} {et~al.}(2013){Hacar}, {Tafalla}, {Kauffmann}, \&
  {Kov{\'a}cs}}]{HacarEtAl2013}
{Hacar}, A., {Tafalla}, M., {Kauffmann}, J., \& {Kov{\'a}cs}, A. 2013, \aap,
  554, A55

\bibitem[{{Heitsch}(2013{\natexlab{a}})}]{Heitsch2013a}
{Heitsch}, F. 2013{\natexlab{a}}, \apj, 769, 115

\bibitem[{{Heitsch}(2013{\natexlab{b}})}]{Heitsch2013b}
---. 2013{\natexlab{b}}, \apj, 776, 62

\bibitem[{{Hennebelle}(2013)}]{Hennebelle2013}
{Hennebelle}, P. 2013, \aap, 556, A153

\bibitem[{{Hennebelle} \& {Andr{\'e}}(2013)}]{HennebelleAndre2013}
{Hennebelle}, P., \& {Andr{\'e}}, P. 2013, \aap, 560, A68

\bibitem[{{Hennebelle} \& {Falgarone}(2012)}]{HennebelleFalgarone2012}
{Hennebelle}, P., \& {Falgarone}, E. 2012, \aapr, 20, 55

\bibitem[{{Hernandez} \& {Tan}(2015)}]{HernandezTan2015}
{Hernandez}, A.~K., \& {Tan}, J.~C. 2015, \apj, 809, 154

\bibitem[{{Heyer} {et~al.}(2009){Heyer}, {Krawczyk}, {Duval}, \&
  {Jackson}}]{HeyerEtAl2009}
{Heyer}, M., {Krawczyk}, C., {Duval}, J., \& {Jackson}, J.~M. 2009, \apj, 699,
  1092

\bibitem[{{Heyer} \& {Brunt}(2004)}]{HeyerBrunt2004}
{Heyer}, M.~H., \& {Brunt}, C.~M. 2004, \apjl, 615, L45

\bibitem[{{Heyer} {et~al.}(2006){Heyer}, {Williams}, \&
  {Brunt}}]{HeyerWilliamsBrunt2006}
{Heyer}, M.~H., {Williams}, J.~P., \& {Brunt}, C.~M. 2006, \apj, 643, 956

\bibitem[{{Hill} {et~al.}(2011){Hill}, {Motte}, {Didelon}, {Bontemps},
  {Minier}, {Hennemann}, {Schneider}, {Andr{\'e}}, {Men'shchikov}, {Anderson},
  {Arzoumanian}, {Bernard}, {di Francesco}, {Elia}, {Giannini}, {Griffin},
  {K{\"o}nyves}, {Kirk}, {Marston}, {Martin}, {Molinari}, {Nguyen Luong},
  {Peretto}, {Pezzuto}, {Roussel}, {Sauvage}, {Sousbie}, {Testi},
  {Ward-Thompson}, {White}, {Wilson}, \& {Zavagno}}]{HillEtAl2011}
{Hill}, T., {Motte}, F., {Didelon}, P., {et~al.} 2011, \aap, 533, A94

\bibitem[{{Hopkins}(2013)}]{Hopkins2013IMF}
{Hopkins}, P.~F. 2013, \mnras, 430, 1653

\bibitem[{{Juvela} {et~al.}(2012{\natexlab{a}}){Juvela}, {Malinen}, \&
  {Lunttila}}]{JuvelaEtAl2012b}
{Juvela}, M., {Malinen}, J., \& {Lunttila}, T. 2012{\natexlab{a}}, \aap, 544,
  A141

\bibitem[{{Juvela} {et~al.}(2012{\natexlab{b}}){Juvela}, {Ristorcelli},
  {Pagani}, {Doi}, {Pelkonen}, {Marshall}, {Bernard}, {Falgarone}, {Malinen},
  {Marton}, {McGehee}, {Montier}, {Motte}, {Paladini}, {T{\'o}th}, {Ysard},
  {Zahorecz}, \& {Zavagno}}]{JuvelaEtAl2012a}
{Juvela}, M., {Ristorcelli}, I., {Pagani}, L., {et~al.} 2012{\natexlab{b}},
  \aap, 541, A12

\bibitem[{{Kainulainen} {et~al.}(2013){Kainulainen}, {Federrath}, \&
  {Henning}}]{KainulainenFederrathHenning2013}
{Kainulainen}, J., {Federrath}, C., \& {Henning}, T. 2013, \aap, 553, L8

\bibitem[{{Kainulainen} {et~al.}(2015){Kainulainen}, {Hacar}, {Alves},
  {Beuther}, {Bouy}, \& {Tafalla}}]{KainulainenEtAl2015}
{Kainulainen}, J., {Hacar}, A., {Alves}, J., {et~al.} 2015, \aap, accepted
  (arXiv:1507.03742)

\bibitem[{{Kauffmann} {et~al.}(2013){Kauffmann}, {Pillai}, \&
  {Goldsmith}}]{KauffmannEtAl2013}
{Kauffmann}, J., {Pillai}, T., \& {Goldsmith}, P.~F. 2013, \apj, 779, 185

\bibitem[{{Kirk} {et~al.}(2015){Kirk}, {Klassen}, {Pudritz}, \&
  {Pillsworth}}]{KirkEtAl2015}
{Kirk}, H., {Klassen}, M., {Pudritz}, R., \& {Pillsworth}, S. 2015, \apj, 802,
  75

\bibitem[{{Kolmogorov}(1941)}]{Kolmogorov1941c}
{Kolmogorov}, A.~N. 1941, Dokl. Akad. Nauk SSSR, 32, 16

\bibitem[{{K{\"o}nyves} {et~al.}(2015){K{\"o}nyves}, {Andr{\'e}},
  {Men'shchikov}, {Palmeirim}, {Arzoumanian}, {Schneider}, {Roy}, {Didelon},
  {Maury}, {Shimajiri}, {Di Francesco}, {Bontemps}, {Peretto}, {Benedettini},
  {Bernard}, {Elia}, {Griffin}, {Hill}, {Kirk}, {Ladjelate}, {Marsh}, {Martin},
  {Motte}, {Nguy{\^e}n Luong}, {Pezzuto}, {Roussel}, {Rygl}, {Sadavoy},
  {Schisano}, {Spinoglio}, {Ward-Thompson}, \& {White}}]{KonyvesEtAl2015}
{K{\"o}nyves}, V., {Andr{\'e}}, P., {Men'shchikov}, A., {et~al.} 2015, \aap,
  584, A91

\bibitem[{{Kritsuk} {et~al.}(2007){Kritsuk}, {Norman}, {Padoan}, \&
  {Wagner}}]{KritsukEtAl2007}
{Kritsuk}, A.~G., {Norman}, M.~L., {Padoan}, P., \& {Wagner}, R. 2007, \apj,
  665, 416

\bibitem[{{Krumholz}(2014)}]{Krumholz2014}
{Krumholz}, M.~R. 2014, Physics Reports, 539, 49

\bibitem[{{Krumholz} {et~al.}(2014){Krumholz}, {Bate}, {Arce}, {Dale},
  {Gutermuth}, {Klein}, {Li}, {Nakamura}, \& {Zhang}}]{KrumholzEtAl2014}
{Krumholz}, M.~R., {Bate}, M.~R., {Arce}, H.~G., {et~al.} 2014, Protostars and
  Planets VI, 243

\bibitem[{{Larson}(1981)}]{Larson1981}
{Larson}, R.~B. 1981, \mnras, 194, 809

\bibitem[{{Mac Low} \& {Klessen}(2004)}]{MacLowKlessen2004}
{Mac Low}, M.-M., \& {Klessen}, R.~S. 2004, RvMP, 76, 125

\bibitem[{{Mac Low} {et~al.}(1998){Mac Low}, {Klessen}, {Burkert}, \&
  {Smith}}]{MacLowEtAl1998}
{Mac Low}, M.-M., {Klessen}, R.~S., {Burkert}, A., \& {Smith}, M.~D. 1998,
  PhRvL, 80, 2754

\bibitem[{{Malinen} {et~al.}(2012){Malinen}, {Juvela}, {Rawlings},
  {Ward-Thompson}, {Palmeirim}, \& {Andr{\'e}}}]{MalinenEtAl2012}
{Malinen}, J., {Juvela}, M., {Rawlings}, M.~G., {et~al.} 2012, \aap, 544, A50

\bibitem[{{McKee} \& {Ostriker}(2007)}]{McKeeOstriker2007}
{McKee}, C.~F., \& {Ostriker}, E.~C. 2007, \araa, 45, 565

\bibitem[{{Men'shchikov} {et~al.}(2010){Men'shchikov}, {Andr{\'e}}, {Didelon},
  {K{\"o}nyves}, {Schneider}, {Motte}, {Bontemps}, {Arzoumanian}, {Attard},
  {Abergel}, {Baluteau}, {Bernard}, {Cambr{\'e}sy}, {Cox}, {di Francesco}, {di
  Giorgio}, {Griffin}, {Hargrave}, {Huang}, {Kirk}, {Li}, {Martin}, {Minier},
  {Miville-Desch{\^e}nes}, {Molinari}, {Olofsson}, {Pezzuto}, {Roussel},
  {Russeil}, {Saraceno}, {Sauvage}, {Sibthorpe}, {Spinoglio}, {Testi},
  {Ward-Thompson}, {White}, {Wilson}, {Woodcraft}, \&
  {Zavagno}}]{MenshchikovEtAl2010}
{Men'shchikov}, A., {Andr{\'e}}, P., {Didelon}, P., {et~al.} 2010, \aap, 518,
  L103

\bibitem[{{Miville-Desch{\^e}nes} {et~al.}(2010){Miville-Desch{\^e}nes},
  {Martin}, {Abergel}, {Bernard}, {Boulanger}, {Lagache}, {Anderson},
  {Andr{\'e}}, {Arab}, {Baluteau}, {Blagrave}, {Bontemps}, {Cohen},
  {Compiegne}, {Cox}, {Dartois}, {Davis}, {Emery}, {Fulton}, {Gry}, {Habart},
  {Huang}, {Joblin}, {Jones}, {Kirk}, {Lim}, {Madden}, {Makiwa}, {Menshchikov},
  {Molinari}, {Moseley}, {Motte}, {Naylor}, {Okumura}, {Pinheiro Gon{\c
  c}alves}, {Polehampton}, {Rod{\'o}n}, {Russeil}, {Saraceno}, {Schneider},
  {Sidher}, {Spencer}, {Swinyard}, {Ward-Thompson}, {White}, \&
  {Zavagno}}]{MivilleDeschenesEtAl2010}
{Miville-Desch{\^e}nes}, M.-A., {Martin}, P.~G., {Abergel}, A., {et~al.} 2010,
  \aap, 518, L104

\bibitem[{{Moeckel} \& {Burkert}(2015)}]{MoeckelBurkert2015}
{Moeckel}, N., \& {Burkert}, A. 2015, \apj, 807, 67

\bibitem[{{Myers}(2011)}]{Myers2011}
{Myers}, P.~C. 2011, \apj, 735, 82

\bibitem[{{Omukai} {et~al.}(2005){Omukai}, {Tsuribe}, {Schneider}, \&
  {Ferrara}}]{OmukaiEtAl2005}
{Omukai}, K., {Tsuribe}, T., {Schneider}, R., \& {Ferrara}, A. 2005, \apj, 626,
  627

\bibitem[{{Ossenkopf} \& {Mac Low}(2002)}]{OssenkopfMacLow2002}
{Ossenkopf}, V., \& {Mac Low}, M.-M. 2002, \aap, 390, 307

\bibitem[{{Ostriker}(1964)}]{Ostriker1964}
{Ostriker}, J. 1964, \apj, 140, 1056

\bibitem[{{Padoan} {et~al.}(2014){Padoan}, {Federrath}, {Chabrier}, {Evans},
  {Johnstone}, {J{\o}rgensen}, {McKee}, \& {Nordlund}}]{PadoanEtAl2014}
{Padoan}, P., {Federrath}, C., {Chabrier}, G., {et~al.} 2014, Protostars and
  Planets VI, 77

\bibitem[{{Palmeirim} {et~al.}(2013){Palmeirim}, {Andr{\'e}}, {Kirk},
  {Ward-Thompson}, {Arzoumanian}, {K{\"o}nyves}, {Didelon}, {Schneider},
  {Benedettini}, {Bontemps}, {Di Francesco}, {Elia}, {Griffin}, {Hennemann},
  {Hill}, {Martin}, {Men'shchikov}, {Molinari}, {Motte}, {Nguyen Luong},
  {Nutter}, {Peretto}, {Pezzuto}, {Roy}, {Rygl}, {Spinoglio}, \&
  {White}}]{PalmeirimEtAl2013}
{Palmeirim}, P., {Andr{\'e}}, P., {Kirk}, J., {et~al.} 2013, \aap, 550, A38

\bibitem[{{Panopoulou} {et~al.}(2014){Panopoulou}, {Tassis}, {Goldsmith}, \&
  {Heyer}}]{PanopoulouEtAl2014}
{Panopoulou}, G.~V., {Tassis}, K., {Goldsmith}, P.~F., \& {Heyer}, M.~H. 2014,
  \mnras, 444, 2507

\bibitem[{{Pillai} {et~al.}(2015){Pillai}, {Kauffmann}, {Tan}, {Goldsmith},
  {Carey}, \& {Menten}}]{PillaiEtAl2015}
{Pillai}, T., {Kauffmann}, J., {Tan}, J.~C., {et~al.} 2015, \apj, 799, 74

\bibitem[{{Pineda} {et~al.}(2010){Pineda}, {Goodman}, {Arce}, {Caselli},
  {Foster}, {Myers}, \& {Rosolowsky}}]{PinedaEtAl2010}
{Pineda}, J.~E., {Goodman}, A.~A., {Arce}, H.~G., {et~al.} 2010, \apjl, 712,
  L116

\bibitem[{{Planck Collaboration} {et~al.}(2014){Planck Collaboration}, {Adam},
  {Ade}, {Aghanim}, {Alves}, {Arnaud}, {Arzoumanian}, {Ashdown}, {Aumont},
  {Baccigalupi}, {Banday}, {Barreiro}, {Bartolo}, {Battaner}, {Benabed},
  {Benoit-L{\'e}vy}, {Bernard}, {Bersanelli}, {Bielewicz}, {Bonaldi},
  {Bonavera}, {Bond}, {Borrill}, {Bouchet}, {Boulanger}, {Bracco}, {Burigana},
  {Butler}, {Calabrese}, {Cardoso}, {Catalano}, {Chamballu}, {Chiang},
  {Christensen}, {Colombi}, {Colombo}, {Combet}, {Couchot}, {Crill}, {Curto},
  {Cuttaia}, {Danese}, {Davies}, {Davis}, {de Bernardis}, {de Rosa}, {de
  Zotti}, {Delabrouille}, {Dickinson}, {Diego}, {Dole}, {Donzelli}, {Dor{\'e}},
  {Douspis}, {Ducout}, {Dupac}, {Efstathiou}, {Elsner}, {En{\ss}lin},
  {Eriksen}, {Falgarone}, {Ferri{\`e}re}, {Finelli}, {Forni}, {Frailis},
  {Fraisse}, {Franceschi}, {Frejsel}, {Galeotta}, {Galli}, {Ganga}, {Ghosh},
  {Giard}, {Gjerl{\o}w}, {Gonz{\'a}lez-Nuevo}, {G{\'o}rski}, {Gregorio},
  {Gruppuso}, {Guillet}, {Hansen}, {Hanson}, {Harrison}, {Henrot-Versill{\'e}},
  {Hern{\'a}ndez-Monteagudo}, {Herranz}, {Hildebrandt}, {Hivon}, {Holmes},
  {Hovest}, {Huffenberger}, {Hurier}, {Jaffe}, {Jaffe}, {Jones},
  {Keih{\"a}nen}, {Keskitalo}, {Kisner}, {Kneissl}, {Knoche}, {Kunz},
  {Kurki-Suonio}, {Lagache}, {Lamarre}, {Lasenby}, {Lattanzi}, {Lawrence},
  {Leonardi}, {Levrier}, {Liguori}, {Lilje}, {Linden-V{\o}rnle},
  {L{\'o}pez-Caniego}, {Lubin}, {Mac{\'{\i}}as-P{\'e}rez}, {Maffei}, {Maino},
  {Mandolesi}, {Maris}, {Marshall}, {Martin}, {Mart{\'{\i}}nez-Gonz{\'a}lez},
  {Masi}, {Matarrese}, {Mazzotta}, {Melchiorri}, {Mendes}, {Mennella},
  {Migliaccio}, {Miville-Desch{\^e}nes}, {Moneti}, {Montier}, {Morgante},
  {Mortlock}, {Munshi}, {Murphy}, {Naselsky}, {Natoli}, {N{\o}rgaard-Nielsen},
  {Noviello}, {Novikov}, {Novikov}, {Oppermann}, {Oxborrow}, {Pagano}, {Pajot},
  {Paoletti}, {Pasian}, {Perdereau}, {Perotto}, {Perrotta}, {Pettorino},
  {Piacentini}, {Piat}, {Plaszczynski}, {Pointecouteau}, {Polenta}, {Ponthieu},
  {Popa}, {Pratt}, {Prunet}, {Puget}, {Rachen}, {Reach}, {Reinecke},
  {Remazeilles}, {Renault}, {Ristorcelli}, {Rocha}, {Roudier},
  {Rubi{\~n}o-Mart{\'{\i}}n}, {Rusholme}, {Sandri}, {Santos}, {Savini},
  {Scott}, {Soler}, {Spencer}, {Stolyarov}, {Sudiwala}, {Sunyaev}, {Sutton},
  {Suur-Uski}, {Sygnet}, {Tauber}, {Terenzi}, {Toffolatti}, {Tomasi},
  {Tristram}, {Tucci}, {Umana}, {Valenziano}, {Valiviita}, {Van Tent},
  {Vielva}, {Villa}, {Wade}, {Wandelt}, {Wehus}, {Wiesemeyer}, {Yvon},
  {Zacchei}, \& {Zonca}}]{PlanckMagneticFilaments2014}
{Planck Collaboration}, {Adam}, R., {Ade}, P.~A.~R., {et~al.} 2014, \aap,
  submitted (arXiv:1409.6728)

\bibitem[{{Planck Collaboration} {et~al.}(2015{\natexlab{a}}){Planck
  Collaboration}, {Ade}, {Aghanim}, {Alves}, {Arnaud}, {Arzoumanian},
  {Ashdown}, {Aumont}, {Baccigalupi}, {Banday}, {Barreiro}, {Bartolo},
  {Battaner}, {Benabed}, {Beno{\^i}t}, {Benoit-L{\'e}vy}, {Bernard},
  {Bersanelli}, {Bielewicz}, {Bock}, {Bonavera}, {Bond}, {Borrill}, {Bouchet},
  {Boulanger}, {Bracco}, {Burigana}, {Calabrese}, {Cardoso}, {Catalano},
  {Chiang}, {Christensen}, {Colombo}, {Combet}, {Couchot}, {Crill}, {Curto},
  {Cuttaia}, {Danese}, {Davies}, {Davis}, {de Bernardis}, {de Rosa}, {de
  Zotti}, {Delabrouille}, {Dickinson}, {Diego}, {Dole}, {Donzelli}, {Dor{\'e}},
  {Douspis}, {Ducout}, {Dupac}, {Efstathiou}, {Elsner}, {En{\ss}lin},
  {Eriksen}, {Falgarone}, {Ferri{\`e}re}, {Finelli}, {Forni}, {Frailis},
  {Fraisse}, {Franceschi}, {Frejsel}, {Galeotta}, {Galli}, {Ganga}, {Ghosh},
  {Giard}, {Gjerl{\o}w}, {Gonz{\'a}lez-Nuevo}, {G{\'o}rski}, {Gregorio},
  {Gruppuso}, {Gudmundsson}, {Guillet}, {Harrison}, {Helou},
  {Henrot-Versill{\'e}}, {Hern{\'a}ndez-Monteagudo}, {Herranz}, {Hildebrandt},
  {Hivon}, {Holmes}, {Hornstrup}, {Huffenberger}, {Hurier}, {Jaffe}, {Jaffe},
  {Jones}, {Juvela}, {Keih{\"a}nen}, {Keskitalo}, {Kisner}, {Knoche}, {Kunz},
  {Kurki-Suonio}, {Lagache}, {Lamarre}, {Lasenby}, {Lattanzi}, {Lawrence},
  {Leonardi}, {Levrier}, {Liguori}, {Lilje}, {Linden-V{\o}rnle},
  {L{\'o}pez-Caniego}, {Lubin}, {Mac{\'{\i}}as-P{\'e}rez}, {Maino},
  {Mandolesi}, {Mangilli}, {Maris}, {Martin}, {Mart{\'{\i}}nez-Gonz{\'a}lez},
  {Masi}, {Matarrese}, {Melchiorri}, {Mendes}, {Mennella}, {Migliaccio},
  {Miville-Desch{\^e}nes}, {Moneti}, {Montier}, {Morgante}, {Mortlock},
  {Munshi}, {Murphy}, {Naselsky}, {Nati}, {Netterfield}, {Noviello}, {Novikov},
  {Novikov}, {Oppermann}, {Oxborrow}, {Pagano}, {Pajot}, {Paladini},
  {Paoletti}, {Pasian}, {Perotto}, {Pettorino}, {Piacentini}, {Piat},
  {Pierpaoli}, {Pietrobon}, {Plaszczynski}, {Pointecouteau}, {Polenta},
  {Ponthieu}, {Pratt}, {Prunet}, {Puget}, {Rachen}, {Reinecke}, {Remazeilles},
  {Renault}, {Renzi}, {Ristorcelli}, {Rocha}, {Rossetti}, {Roudier},
  {Rubi{\~n}o-Mart{\'{\i}}n}, {Rusholme}, {Sandri}, {Santos}, {Savelainen},
  {Savini}, {Scott}, {Soler}, {Stolyarov}, {Sudiwala}, {Sutton}, {Suur-Uski},
  {Sygnet}, {Tauber}, {Terenzi}, {Toffolatti}, {Tomasi}, {Tristram}, {Tucci},
  {Umana}, {Valenziano}, {Valiviita}, {Van Tent}, {Vielva}, {Villa}, {Wade},
  {Wandelt}, {Wehus}, {Ysard}, {Yvon}, \&
  {Zonca}}]{PlanckMagneticFilaments2015_1}
{Planck Collaboration}, {Ade}, P.~A.~R., {Aghanim}, N., {et~al.}
  2015{\natexlab{a}}, \aap, accepted (arXiv:1502.04123)

\bibitem[{{Planck Collaboration} {et~al.}(2015{\natexlab{b}}){Planck
  Collaboration}, {Ade}, {Aghanim}, {Arnaud}, {Ashdown}, {Aumont},
  {Baccigalupi}, {Banday}, {Barreiro}, {Bartolo}, {Battaner}, {Benabed},
  {Benoit-L{\'e}vy}, {Bernard}, {Bersanelli}, {Bielewicz}, {Bonaldi},
  {Bonavera}, {Bond}, {Borrill}, {Bouchet}, {Boulanger}, {Bracco}, {Burigana},
  {Calabrese}, {Cardoso}, {Catalano}, {Chamballu}, {Chary}, {Chiang},
  {Christensen}, {Colombo}, {Combet}, {Crill}, {Curto}, {Cuttaia}, {Danese},
  {Davies}, {Davis}, {de Bernardis}, {de Rosa}, {de Zotti}, {Delabrouille},
  {Delouis}, {Dickinson}, {Diego}, {Dole}, {Donzelli}, {Dor{\'e}}, {Douspis},
  {Dunkley}, {Dupac}, {Efstathiou}, {Elsner}, {En{\ss}lin}, {Eriksen},
  {Falgarone}, {Ferri{\`e}re}, {Finelli}, {Forni}, {Frailis}, {Fraisse},
  {Franceschi}, {Frolov}, {Galeotta}, {Galli}, {Ganga}, {Ghosh}, {Giard},
  {Gjerl{\o}w}, {Gonz{\'a}lez-Nuevo}, {G{\'o}rski}, {Gruppuso}, {Guillet},
  {Hansen}, {Harrison}, {Helou}, {Hern{\'a}ndez-Monteagudo}, {Herranz},
  {Hildebrandt}, {Hivon}, {Hornstrup}, {Hovest}, {Huang}, {Huffenberger},
  {Hurier}, {Jaffe}, {Jones}, {Juvela}, {Keih{\"a}nen}, {Keskitalo}, {Kisner},
  {Kneissl}, {Knoche}, {Kunz}, {Kurki-Suonio}, {Lamarre}, {Lasenby},
  {Lattanzi}, {Lawrence}, {Leonardi}, {Le{\'o}n-Tavares}, {Levrier}, {Liguori},
  {Lilje}, {Linden-V{\o}rnle}, {L{\'o}pez-Caniego}, {Lubin},
  {Mac{\'{\i}}as-P{\'e}rez}, {Maffei}, {Maino}, {Mandolesi}, {Maris}, {Martin},
  {Mart{\'{\i}}nez-Gonz{\'a}lez}, {Masi}, {Matarrese}, {McGehee}, {Melchiorri},
  {Mennella}, {Migliaccio}, {Miville-Desch{\^e}nes}, {Moneti}, {Montier},
  {Morgante}, {Mortlock}, {Munshi}, {Murphy}, {Naselsky}, {Nati}, {Natoli},
  {Novikov}, {Novikov}, {Oppermann}, {Oxborrow}, {Pagano}, {Pajot}, {Paoletti},
  {Pasian}, {Perdereau}, {Pettorino}, {Piacentini}, {Piat}, {Pierpaoli},
  {Plaszczynski}, {Pointecouteau}, {Polenta}, {Ponthieu}, {Pratt}, {Prunet},
  {Puget}, {Rachen}, {Reach}, {Rebolo}, {Reinecke}, {Remazeilles}, {Renault},
  {Renzi}, {Ristorcelli}, {Rocha}, {Rosset}, {Rossetti}, {Roudier},
  {Rubi{\~n}o-Mart{\'{\i}}n}, {Rusholme}, {Sandri}, {Santos}, {Savelainen},
  {Savini}, {Scott}, {Serra}, {Soler}, {Stolyarov}, {Sudiwala}, {Sunyaev},
  {Suur-Uski}, {Sygnet}, {Tauber}, {Terenzi}, {Toffolatti}, {Tomasi},
  {Tristram}, {Tucci}, {Umana}, {Valenziano}, {Valiviita}, {Van Tent},
  {Vielva}, {Villa}, {Wade}, {Wandelt}, {Wehus}, {Yvon}, {Zacchei}, \&
  {Zonca}}]{PlanckMagneticFilaments2015_2}
---. 2015{\natexlab{b}}, \aap, accepted (arXiv:1505.02779)

\bibitem[{{Price} {et~al.}(2011){Price}, {Federrath}, \&
  {Brunt}}]{PriceFederrathBrunt2011}
{Price}, D.~J., {Federrath}, C., \& {Brunt}, C.~M. 2011, \apjl, 727, L21

\bibitem[{{Ricker}(2008)}]{Ricker2008}
{Ricker}, P.~M. 2008, \apjs, 176, 293

\bibitem[{{Roman-Duval} {et~al.}(2011){Roman-Duval}, {Federrath}, {Brunt},
  {Heyer}, {Jackson}, \& {Klessen}}]{RomanDuvalEtAl2011}
{Roman-Duval}, J., {Federrath}, C., {Brunt}, C., {et~al.} 2011, \apj, 740, 120

\bibitem[{{Roy} {et~al.}(2015){Roy}, {Andre'}, {Arzoumanian}, {Peretto},
  {Palmeirim}, {Konyves}, {Schneider}, {Benedettini}, {Di Francesco}, {Elia},
  {Hill}, {Ladjelate}, {Louvet}, {Motte}, {Pezzuto}, {Schisano}, {Shimajiri},
  {Spinoglio}, {Ward-Thompson}, \& {White}}]{RoyEtAl2015}
{Roy}, A., {Andre'}, P., {Arzoumanian}, D., {et~al.} 2015, \aap, accepted
  (arXiv:1509.01819)

\bibitem[{{Schmidt} {et~al.}(2009){Schmidt}, {Federrath}, {Hupp}, {Kern}, \&
  {Niemeyer}}]{SchmidtEtAl2009}
{Schmidt}, W., {Federrath}, C., {Hupp}, M., {Kern}, S., \& {Niemeyer}, J.~C.
  2009, \aap, 494, 127

\bibitem[{{Schmidt} {et~al.}(2006){Schmidt}, {Hillebrandt}, \&
  {Niemeyer}}]{SchmidtHillebrandtNiemeyer2006}
{Schmidt}, W., {Hillebrandt}, W., \& {Niemeyer}, J.~C. 2006, CF, 35, 353

\bibitem[{{Schneider} {et~al.}(2012){Schneider}, {Csengeri}, {Hennemann},
  {Motte}, {Didelon}, {Federrath}, {Bontemps}, {Di Francesco}, {Arzoumanian},
  {Minier}, {Andr{\'e}}, {Hill}, {Zavagno}, {Nguyen-Luong}, {Attard},
  {Bernard}, {Elia}, {Fallscheer}, {Griffin}, {Kirk}, {Klessen}, {K{\"o}nyves},
  {Martin}, {Men'shchikov}, {Palmeirim}, {Peretto}, {Pestalozzi}, {Russeil},
  {Sadavoy}, {Sousbie}, {Testi}, {Tremblin}, {Ward-Thompson}, \&
  {White}}]{SchneiderEtAl2012}
{Schneider}, N., {Csengeri}, T., {Hennemann}, M., {et~al.} 2012, \aap, 540, L11

\bibitem[{{Schneider} \& {Elmegreen}(1979)}]{SchneiderElmegreen1979}
{Schneider}, S., \& {Elmegreen}, B.~G. 1979, \apjs, 41, 87

\bibitem[{{Seifried} \& {Walch}(2015)}]{SeifriedWalch2015}
{Seifried}, D., \& {Walch}, S. 2015, \mnras, 452, 2410

\bibitem[{{Shetty} {et~al.}(2012){Shetty}, {Beaumont}, {Burton}, {Kelly}, \&
  {Klessen}}]{ShettyEtAl2012}
{Shetty}, R., {Beaumont}, C.~N., {Burton}, M.~G., {Kelly}, B.~C., \& {Klessen},
  R.~S. 2012, \mnras, 425, 720

\bibitem[{{Smith} {et~al.}(2014){Smith}, {Glover}, \&
  {Klessen}}]{SmithGloverKlessen2014}
{Smith}, R.~J., {Glover}, S.~C.~O., \& {Klessen}, R.~S. 2014, \mnras, 445, 2900

\bibitem[{{Smith} {et~al.}(2016){Smith}, {Glover}, {Klessen}, \&
  {Fuller}}]{SmithEtAl2016}
{Smith}, R.~J., {Glover}, S.~C.~O., {Klessen}, R.~S., \& {Fuller}, G.~A. 2016,
  \mnras, 455, 3640

\bibitem[{{Solomon} {et~al.}(1987){Solomon}, {Rivolo}, {Barrett}, \&
  {Yahil}}]{SolomonEtAl1987}
{Solomon}, P.~M., {Rivolo}, A.~R., {Barrett}, J., \& {Yahil}, A. 1987, \apj,
  319, 730

\bibitem[{{Sousbie}(2011)}]{Sousbie2011}
{Sousbie}, T. 2011, \mnras, 414, 350

\bibitem[{{Sousbie} {et~al.}(2011){Sousbie}, {Pichon}, \&
  {Kawahara}}]{SousbieEtAl2011}
{Sousbie}, T., {Pichon}, C., \& {Kawahara}, H. 2011, \mnras, 414, 384

\bibitem[{{Stone} {et~al.}(1998){Stone}, {Ostriker}, \&
  {Gammie}}]{StoneOstrikerGammie1998}
{Stone}, J.~M., {Ostriker}, E.~C., \& {Gammie}, C.~F. 1998, \apjl, 508, L99

\bibitem[{{Sugitani} {et~al.}(2011){Sugitani}, {Nakamura}, {Watanabe},
  {Tamura}, {Nishiyama}, {Nagayama}, {Kandori}, {Nagata}, {Sato}, {Gutermuth},
  {Wilson}, \& {Kawabe}}]{SugitaniEtAl2011}
{Sugitani}, K., {Nakamura}, F., {Watanabe}, M., {et~al.} 2011, \apj, 734, 63

\bibitem[{{Sur} {et~al.}(2010){Sur}, {Schlei\-cher}, {Banerjee}, {Federrath},
  \& {Klessen}}]{SurEtAl2010}
{Sur}, S., {Schlei\-cher}, D.~R.~G., {Banerjee}, R., {Federrath}, C., \&
  {Klessen}, R.~S. 2010, \apjl, 721, L134

\bibitem[{{Tomisaka}(2014)}]{Tomisaka2014}
{Tomisaka}, K. 2014, \apj, 785, 24

\bibitem[{{Tremblay} {et~al.}(2015){Tremblay}, {O'Dea}, {Baum}, {Mittal},
  {McDonald}, {Combes}, {Li}, {McNamara}, {Bremer}, {Clarke}, {Donahue},
  {Edge}, {Fabian}, {Hamer}, {Hogan}, {Oonk}, {Quillen}, {Sanders},
  {Salom{\'e}}, \& {Voit}}]{TremblayEtAl2015}
{Tremblay}, G.~R., {O'Dea}, C.~P., {Baum}, S.~A., {et~al.} 2015, \mnras, 451,
  3768

\bibitem[{{Truelove} {et~al.}(1997){Truelove}, {Klein}, {McKee}, {Holliman},
  {Howell}, \& {Greenough}}]{TrueloveEtAl1997}
{Truelove}, J.~K., {Klein}, R.~I., {McKee}, C.~F., {et~al.} 1997, \apjl, 489,
  L179

\bibitem[{{V{\'a}zquez-Semadeni}(1994)}]{Vazquez1994}
{V{\'a}zquez-Semadeni}, E. 1994, \apj, 423, 681

\bibitem[{{V{\'a}zquez-Semadeni} {et~al.}(2003){V{\'a}zquez-Semadeni},
  {Ballesteros-Paredes}, \& {Klessen}}]{VazquezBallesterosKlessen2003}
{V{\'a}zquez-Semadeni}, E., {Ballesteros-Paredes}, J., \& {Klessen}, R.~S.
  2003, \apjl, 585, L131

\bibitem[{{Waagan} {et~al.}(2011){Waagan}, {Federrath}, \&
  {Klingenberg}}]{WaaganFederrathKlingenberg2011}
{Waagan}, K., {Federrath}, C., \& {Klingenberg}, C. 2011, Journal of
  Computational Physics, 230, 3331

\bibitem[{{Wang} {et~al.}(2015){Wang}, {Testi}, {Ginsburg}, {Walmsley},
  {Molinari}, \& {Schisano}}]{WangEtAl2015}
{Wang}, K., {Testi}, L., {Ginsburg}, A., {et~al.} 2015, \mnras, 450, 4043

\bibitem[{{Wang} {et~al.}(2010){Wang}, {Li}, {Abel}, \&
  {Nakamura}}]{WangEtAl2010}
{Wang}, P., {Li}, Z.-Y., {Abel}, T., \& {Nakamura}, F. 2010, \apj, 709, 27

\end{thebibliography}

\end{document}